\documentclass[superscriptaddress,secnumarabic,
amssymb,amsmath,nobibnotes,aps,prd,showkeys,showpacs,nofootinbib]{revtex4}%
\usepackage{graphicx}
\usepackage{epsf}
\usepackage{bm}
\usepackage{amsmath}
\usepackage{amsfonts}
\usepackage{amssymb}
\usepackage{epstopdf}
\usepackage{subfigure}
\usepackage{natbib}
\usepackage{color}%
\usepackage{hyperref}
\hypersetup{dvips,dvipdfm,linktoc=page,colorlinks=true,linkcolor=blue,citecolor=red,filecolor=magenta,urlcolor=magenta,bookmarks=true}
\providecommand{\U}[1]{\protect\rule{.1in}{.1in}}
\newcommand{\be}{\begin{equation}}
	\newcommand{\ee}{\end{equation}}

\newcommand{\mincir}{\raise
	-3.truept\hbox{\rlap{\hbox{$\sim$}}\raise4.truept\hbox{$<$}\ }}
\newcommand{\magcir}{\raise
	-3.truept\hbox{\rlap{\hbox{$\sim$}}\raise4.truept\hbox{$>$}\ }}

\begin{document}
	\title{Phase space analysis and cosmography of a two-fluid cosmological model}
	\author{Goutam Mandal}
	\email{gmandal243@gmail.com; 
		rs\_goutamm@nbu.ac.in}
	\affiliation{Department of Mathematics, University of North Bengal, Raja Rammohunpur, Darjeeling-734013, West Bengal, India.}
	\author{Sujay Kr. Biswas\footnote{corresponding author}}
	\email{sujaymathju@gmail.com; sujay.math@nbu.ac.in}
	\affiliation{Department of Mathematics, University of North Bengal, Raja Rammohunpur, Darjeeling-734013, West Bengal, India.}
\keywords{Modified chaplygin gas, Tachyon field, Phase space, stability, Cosmography}
	\pacs{95.36.+x, 95.35.+d, 98.80.-k, 98.80.Cq.}
\begin{abstract}
In the framework of spatially flat Friedmann-Lemaitre-Robertson-Walker (FLRW) space-time, we investigate a two-fluid cosmological model where a tachyon  scalar field with self-interacting potential and a modified chaplygin gas with non-linear equation of state are taken as the background fluids. We perform phase space analysis of the autonomous system obtained from the cosmological governing equations by a suitable transformation of variables. Linear stability theory is employed to characterise the stability criteria for hyperbolic critical points. Numerical investigation is carried out for non-hyperbolic points. Our study reveals that modified chaplygin fluid dominated solutions cannot provide the late-time evolution. Late-time accelerated evolution is obtained only when the solution is dominated by tachyon fluid. This study also yields a late-time scaling attractor providing similar order of energy densities in its evolution. The adiabatic sound speed is evaluated for both the fluids and test the stability of the models independently. Further, we perform cosmographic analysis in the model independent way by evaluating all the cosmographic parameters and then $Om$ diagnostic is also found to compare our model with $\Lambda$CDM model.
\end{abstract}

\maketitle
\section{Introduction}
One of the main puzzles in recent theoretical cosmology is the accelerated expansion of the universe \cite{Riess:1998,Perlmutter:1998}. This phenomenon cannot be explained by considering that the universe is only filled with normal matters. Then, in one approach within the general relativity, it is required to postulate an exotic type of fluid that dominates the present epoch of the universe. The mysterious fluid with a huge negative pressure and violating the strong energy condition (i.e., $\rho+ 3p<0$), is dubbed as dark energy (DE). A very little is known about DE, it has only theoretical existence. However, the gravitational repulsive force driven by negative pressure ($p<-\frac{\rho}{3}$) by the vacuum energy is responsible to accelerate the expansion of the universe. Many DE models have been studied in the literature, see for instance \cite{Copeland 2006,Bamba 2012}. 

It is assumed that the cosmological constant ($\Lambda$) is the most preferred DE candidate till now. The $\Lambda$ was first introduced by Einstein himself in his field equation to obtain a static universe. However, after Hubble's observation, cosmologists have taken up the cosmological constant as a cosmic fluid and it can play a role of DE candidate which with its constant negative Equation of State (EoS) parameter: $\omega_{\Lambda}=\frac{p_{\Lambda}}{\rho_{\Lambda}}=-1$ drives the present acceleration of the universe. Historically, Georges Lemaître in 1934 proposed the modern interpretation of $\Lambda$ as representing the energy density of `Lorentz Invariant Vacuum Energy' (LIVE). In an adiabatic expansion of homogeneous universe the Friedmann equations indicate that the density of the LIVE is constant during the expansion and hence this can be represented by the cosmological constant. A detail analysis about the LIVE can be found in an interesting work in Ref.\cite{Gron 2018} and the references therein. The $\Lambda$ together with cold-dark-matter (CDM) constitutes the famous $\Lambda$CDM model that, according to several observations, provides the best fit results to the observational data till now \cite{Carroll 1992,Peebles 2003,Sahni 2000,Planck 2016}. However, the model has two theoretical issues, namely, cosmological constant problem \cite{Weinberg 1989,Sahni 2000,Padmanabhan 2003} and coincidence problem \cite{Zlatev 1999}. 
In a well established approach, various dark energy models have been widely studied in literature \cite{Frieman 2008,Li 2011} where the cosmological constant replaces another fluid whose equation of state is non-constant. It varies with time and it depends on variety of extra parameters also \cite{Cooray 1999,Linder 2003,Sahni 2003,Padmanabhan 2003,Wetterich 2004,Komatsu 2009,Wang 2008,Sendra 2012}. Thus one can formulate a background extra fluid model (alternative to $\Lambda$CDM) that satisfies the two energy limits dominating at low redshift and at high redshift and preserves the expansion history of the universe (an early matter-dominated era and an accelerated expansion at late phase). One of such models can be formulated by a single perfect fluid model with an exotic equation of state $p=-\frac{\beta}{\rho}$ where $\beta$ is positive constant and the model belongs to the so-called famous Chaplygin gas fluid. It has interesting feature that it can describe an unified dark energy-dark matter problem within a single model description \cite{Kamenshchik 2001}. Also, it can be reinterpreted as a solution to an interacting dark sector problem \cite{Santiago 2012}. After that several generalizations of the model have been made in the literature \cite{Lazkoz 2019,Salahedin 2020,Mamon 2021,Khurshudyan 2017,Ranjit 2014,Li 2009}. Also, for a brief review, see the Ref. \cite{Pourhassen 2014}. In this regard, a most popular and promising generalization is the Modified Chaplygin Gas (MCG) with non-linear equation of state ($p=\gamma\rho-\frac{\beta}{\rho^{n}},~~ 0\leq n\leq1,~\gamma\geq 0$). The MCG equation of state shows its capability of providing the radiation era to $\Lambda$CDM within this single model \cite{Debnath 2004,Jamil 2009} and this type of feature has attracted many people to look after the fluid in cosmological study.  \\

In another approach, Quintessence can be treated as candidate for DE model. This is the simplest and popular DE model represented by the canonical scalar field $\phi(t)$ \cite{Copeland 2006,Bamba 2012}. A generalization of the scalar field with Lagrangian can be described as \cite{Padmanabhan 2002} $L=-V(\phi) \sqrt{1-\partial ^{i} \phi \partial_{i} \phi}$ where the $\phi$ is referred to as tachyon field and $V(\phi)$ is the potential of the tachyon field. Inflation could provide a mechanism for production of density perturbations which is required to shape the evolution of the universe. Simply, in an inflationary dynamics, the universe is dominated by scalar field where the potential term dominates over the kinetic term, followed by a reheating period \cite{Gibbons2003, Jassal2004}. However, there is no well accepted interpretation to integrate inflationary scenario in early universe where the scalar field drives inflation with one of the known fields of particle physics. It is also remarkable to emanate inflation potential naturally from the underlying fundamental theory. In this context, the tachyon scalar field resembling to unstable D-branes could be responsible for inflation in the early universe \cite{Sami2004, Nozari_Rashidi2013}. However, in context of string theory as well as in string field theory, the tachyon field arises naturally and could be accountable for inflation in early time. The tachyon field is widely investigated due to its eventual role in DBI (Dirac-Born-Infeld) action which is used to recite the D-brane action \cite{Sen 2002a,Sen 2002b,Sen 2002c,Garousi 2003}. Thus, the effective field of such tachyons can be taken into consideration as DBI scalar field\cite{Sen 2002b}. Since then, the DBI scalar field has been studied (see for instance, in Refs. \cite{Gibbons2003,Padmanabhan 2002a}) in the framework of FLRW cosmology in understanding different cosmic phases of evolution. For example, late-time acceleration can also be realized by the DBI scalar field, see in Ref.\cite{Bhagala 2003}. The DBI has also been proposed (see in Ref. \cite{Padmanabhan 2002}) as an alternative choice of dark matter and it can play an important role in late-time cosmology . Furhermore, another alternative approach has been made in Ref.\cite{Gorini 2003} where DBI scalar field is characterised as the Modified Chaplygin gas. Interestingly, the DBI scalar field is investigated from the dynamical systems perspective (see for instance, in refences \cite{Copeland 2005,Aguirregabiria 2004}). Recently, the authors in Ref.\cite{Ghosh 2024} have studied the DBI scalar field in context of modified gravity in the framework of dynamical systems where different forms of potentials are taken into account.    \\

The authors in Ref.\cite{Kamenshchik 2001} show that the chaplygin gas equation of state can indeed come from a scalar field. Also, the viablity of chaplygin gas in order to explain the quintessence has been investigated in Ref. \cite{Garousi 2003} where it has been shown that the chaplygin gas equation of state can indeed sustain tracker solution which could be used lucrative alternative to dark energy. It can be shown that even from a tachyon like scalar potential chaplygin equation of state can emerge with a particular solution of kinetic and potential term \cite{Benaoum 2022}. Finally, last but not the least it has been shown that chaplygin gas equation of state naturally emerges as a Brane like scenario as discussed in Ref.\cite{Kamenshchik 2000}.

From the above, the cosmological study of the mixture of two-fluid components: modified chaplygin gas and tachyon field has a growing amount of interest nowadays. In this regards, one may refer the references in \cite{Noorbakhsh 2013,Amani 2013,Benaoum 2022} where the authors have investigated the evolution of FRW universe when the universe is assumed to be filled with mixture of modified chaplygin gas and tachyon field. 

In the present work, we are interested to investigate the model of two-fluid cosmology. The modified chaplygin gas  is taken as a background fluid with its non-linear equation of state and the tachyon fluid is considered as a source of DE. First, we perform dynamical analysis to the model. The dynamical system analysis may be considered to be a powerful tool to extract an overall qualitative information from  the cosmological models with dark energy-dark matter problems (see for references \cite{Biswas 2015a,Biswas 2015b,Chen 2009,Odintsov 2018b,Aljaf 2020,Oikonomou 2019,Biswas 2021,Mandal 2022,Bahamonde 2018,Coley 2003,Biswas 2017,Teixeira 2019,Odintsov2017,Oikonomou2018,Odintsov2018a,Kleidis2018,Mandal 2024}). Interested reader may follow the recent article \cite{Ghosh 2024a} for dynamical analysis to scalar field model in modified gravity where exponential as well as powerlaw potential are studied and the work in ref.\cite{Leon 2023} has investigated a unified dynamical system analysis at a scalar field's background and perturbation levels with arbitrary potentials. In a recent work \cite{Ghosh 2024}, the authors havestudied  the DBI scalar field by using dynamical systems analysis.
In our paper, first we convert our cosmological model into an autonomous system of ordinary differential equations by suitable choice of dynamical variables. Then we extract critical points from autonomous system. The cosmological parameters are expressed in terms of dynamical variables. Then, we find the stability of the critical points by evaluating the eigenvalues of linearized Jacobian matrix at critical points. From the phase space of the model we consider, it is observed that critical points describe: (a) Modified chaplygin gas dominated solutions, (b) tachyonic fluid dominated solution and (c) scaling solution.
The MCG fluid dominated solutions can never be able to give late-time solution from the dynamical analysis (because there is no stable solution in phase space). These solutions can mimic the radiation dominated or dust dominated decelerated phase providing the intermediate era. On the other hand, tachyonic fluid dominated solution can represent late-time accelerated evolution in quintessence era. The scaling solution can represent the late-time evolution corresponding to accelerated universe attracted in cosmological constant era solving the coincidence problem too. Numerical investigation of the cosmological evolution of cosmological parameters show that the model mimics the $\Lambda$CDM. Classical stability for the model is also investigated by finding the squared sound speed which shows the model is stable classically.\\

We also test the model by evaluating the geometrical cosmographic parameters $r$ and $s$ which are obtained from the scale factor $a(t)$. We then investigate the discrimination of our model with $\Lambda$CDM by plotting the trajectories in $r-s$ plane and in $r-q$ plane. This shows that our model can mimic the $\Lambda$CDM model successfully. Further, we study the more general geometric parameters, namely, the jerk ($j$), snap ($s$) and the lerk ($l$) parameters. The behaviour of the trajectories show the evolution of our model mimics the $\Lambda$CDM in late-times. $Om$ diagnostic is also studied by finding the Hubble parameter and this shows the model converges to the $\Lambda$CDM in future.  \\

The organization of the paper is as follows:\\    
In the next section \ref{model and autonomous system}, we present the basic equations of the cosmological models with modified chaplygin gas and the tachyon field. Then we construct autonomous system and cosmological parameters in terms of dynamical variables. In section \ref{phase space autonomous system} phase space analysis is performed by finding the local stability criteria of the critical points. Cosmological implications of the points are discussed in section \ref{cosmological implications}. Cosmography analysis of the model is presented in the section \ref{cosmography}. Finally, we conclude with a short discussion in section \ref{discussion}.

\section{ Cosmological model with Modified Chaplygin gas and Tachyonic field and formulation of autonomous system }\label{model and autonomous system}

In this section, we present the cosmological governing equations with modified chaplygin gas and the tachyon field in FLRW cosmological model of two non-interacting fluids. We construct the autonomous system of ordinary differential equations by adopting some dynamical variables. 

\subsection{The cosmological two-fluid model}
The equation of state for Chaplygin gas (single component fluid) is
 $$p=-\frac{\beta}{\rho}.$$
The generalized Chaplygin gas (GCG) is described by the following EoS: 
 $$p=-\frac{\beta}{\rho^{n}},~~ 0\leq n\leq1,~~\beta~~\mbox{is positive constant}.$$ 
The Modified Chaplygin Gas (MCG) is defined in the following non-linear perfect fluid equation of state
\begin{equation}\label{MCG EoS}
	 p=\gamma\rho-\frac{\beta}{\rho^{n}},~~ 0\leq n\leq1,
\end{equation}
where $\gamma\geq0$ and $\beta>0$ are constants.
Therefore, it is clear that for $\beta=0$, one obtains the perfect fluid equation of state $p=\gamma \rho$. However, for $\gamma=0$, the MCG corresponds to GCG and in addition $n=1$, fluid can behave of CG EoS: $p=-\frac{\beta}{\rho}$. Therefore, we can comment that the MCG is the mixture of barotropic perfect fluid and Generalized Chaplygin gas. Note that the MCG exhibits the radiation era in the early times high energy density i.e., when $a$ is small and $\rho\rightarrow\infty$. On the other hand, the MCG can behave as cosmological constant at late times (at low energy density). Also, it worthy to note that the equation of state of LIVE (Lorentz Invariant Vacuum Energy), corresponding to a cosmological constant, is $p=-\rho$. Inserting this in Eqn (\ref{MCG EoS}) gives $\rho=\left(\frac{\beta}{1+\gamma} \right)^{\frac{1}{n+1}}$. This appears as the final density for large values of the scale factor (at late-times).

Thus, depending on energy scale, the MCG has a beauty of nature in describing the dark sectors of the Universe. It behaves as dark matter (DM) at high densities and transforms into dark energy (DE) at lower ones.\\
  
Now, we shall start with the spatially flat, symmetric and homogeneous 4D space-time line element with scale factor $a(t)$:
 \begin{equation}\label{FLRW metric}
 	ds^2=dt^2 -a^{2}(t) \left(dr^2 +r^2 d\Omega^2\right)
  \end{equation} 
The Einstein field equations:
 \begin{equation}\label{EFE}
	R_{j}^{i}-\frac{1}{2} \delta_{j}^{i} R=\kappa^{2} T_{j}^{i}
\end{equation}
The energy-momentum tensor for the fluid content of the universe is:
 \begin{equation}\label{stress-energy tensor}
	 T_{j}^{i}=(\rho+p) u^{i} u_{j} -p \delta_{j}^{i}
\end{equation}
The salient feature of the energy-momentum tensor of the above is that it could be assumed as the sum of two fluids: tachyon and modified chaplygin gas. Here, we explicitly will break up the density $\rho$ and pressure $p$ in Eqn (\ref{stress-energy tensor}) and write them in the following form: $\rho=\rho_{\phi}+\rho_{CG}$, where $\rho_{\phi}$ is the energy density for tachyon field and $\rho_{CG}$ refers to the energy density for the modified chaplygin gas. This means that the energy momentum tensor can be thought as made up of two ingredients. The Einstein field equations (\ref{EFE}) for the above metric (\ref{FLRW metric}) will give the following Friedmann equation and acceleration equation (after considering the natural units $\left(\kappa^{2}=8\pi G=c=1  \right)$)
\begin{equation}\label{Friedmann}
3 H^{2} =\rho_{total}
\end{equation}
and
\begin{equation}\label{Raychaudhuri}
 2 \dot{H}+3H^{2} =-p_{total},
\end{equation}
where the Hubble parameter is expressed in terms of scale factor and its derivative as $H=\frac{\dot{a}}{a}$ which defines the expansion of the universe. An over-dot stands for the differentiation with respect to the cosmic time $t$. We consider the universe is filled with mixture of two fluids: one is tachyon fluid based on scalar field and other is modified Chaplygin gas with perfect fluid equation of state. Total energy density is the sum of energy densities of tachyon plus MCG fluids and is estimated as 
\begin{equation}\label{rho tot}
	\rho_{total}=\rho_{\phi}+\rho_{CG}
\end{equation}
Similarly, the total pressure is the sum of pressure density of tachyon plus the pressure of MCG and is defined as
\begin{equation}\label{p tot}
	p_{total}=p_{\phi}+p_{CG}
\end{equation}
The tachyon field is characterized by its EoS: $\omega_{\phi}= \frac{p_{\phi}}{\rho_{\phi}}$, where the energy density $\rho_{\phi}$ and pressure $p_{\phi}$ of tachyonic field are defined as: 
\begin{equation}\label{energy density scalar}
	\rho_{\phi}=\frac{V(\phi)}{\sqrt{1-\dot{\phi}^{2}}}
\end{equation}\\
and 
\begin{equation}\label{pressure scalar}
	p_{\phi}=-V(\phi)\sqrt{1-\dot{\phi}^{2}},
\end{equation}
where $V(\phi)$ is the potential function of the scalar field and $\dot{\phi}^2$ is for the kinetic part of the scalar field. The equation of state parameter for the MCG can be defined as (in Eqn (\ref{MCG EoS})):
\begin{equation}\label{MCG EoS 1}
	p_{CG}=\gamma~ \rho_{CG} -\frac{\beta}{\rho_{CG} ^{n}},~~ 0\leq n\leq1,
\end{equation}
The total (effective) equation of state parameter for the model can be expressed as
\begin{equation}\label{equations}
	\omega_{total}=\frac{p_{total}}{\rho_{total}}= \frac{p_{\phi}+p_{CG}}{\rho_{\phi}+\rho_{CG}}.
\end{equation}
which defines the evolutionary era of the universe.
Now, since we consider the space-time is filled with the fluid (tachyon $+$ MCG) of energy density $\rho_{total}$ and pressure $p_{total}$, then the energy-momentum conservation ($T_{; i}^{i j}=0$) gives
\begin{equation}\label{continuity total}
	\dot{\rho}_{total}+3H(\rho_{total}+p_{total})=0,
\end{equation}
We consider that the individual components of fluids do not interact and each of them separately conserved which implies that
\begin{equation}\label{continuity}
	\dot{\rho}_\phi+3H(\rho_\phi+p_{\phi})=0,
\end{equation}
and 
\begin{equation}\label{continuity_CG}
	\dot{\rho}_{CG}+3H(\rho_{CG}+p_{CG})=0,
\end{equation}
The governing equations of our model are
\begin{equation}\label{Friedmann1}
	3 H^{2} =\frac{V(\phi)}{\sqrt{1-\dot{\phi}^{2}}}+\rho_{CG}
\end{equation}
and
\begin{equation}\label{Raychaudhuri1}
	2 \dot{H}+3H^{2} =V(\phi)\sqrt{1-\dot{\phi}^{2}}-p_{CG},
\end{equation}
where, $\rho_{CG}$ and $p_{CG}$ are given in Eqn (\ref{MCG EoS 1}).
The evolution equation for the tachyonic scalar field (from Eqn. (\ref{continuity}), using Eqn. (\ref{energy density scalar}) and Eqn. (\ref{pressure scalar})) takes the form 
\begin{equation}\label{EvolutionEqn-scalar}
  \frac{\ddot{\phi}}{1-\dot{\phi}^{2}}+3H\dot{\phi}+\frac{V'(\phi)}{V(\phi)}=0
\end{equation}  
where $'\equiv \frac{d}{d\phi}$ represents the derivative with respect to Tachyonic scalar field $\phi$. 

\subsection{Autonomous system and cosmological parameters}\label{autonomous system and parameters}
We shall now discuss the dynamical analysis of the model by constructing the autonomous system of ordinary differential equations from cosmological evolution equations derived in previous subsection. First, we shall adopt a suitable set of dynamical variables in the following in terms of cosmological variables.  
\begin{equation}\label{dynamical_variables}
    x=\dot{\phi},~~~y=\frac{V(\phi)}{3H^2},~~~z=\frac{\rho_{CG}}{3H^2}~~~\mbox{and}~~~s=\frac{p_{CG}}{3H^2}
\end{equation}\\
By using the above variables in (\ref{dynamical_variables}) in the governing equations (\ref{continuity}), (\ref{continuity_CG}), (\ref{Friedmann1}), (\ref{Raychaudhuri1}), we frame the four dimensional autonomous system of ordinary differential equations:
\begin{eqnarray}
	\begin{split}
		\frac{dx}{dN}& =\left( x^{2}-1\right) \left( 3x+\sqrt{3}\lambda\sqrt{y}\right)  ,& \\
		\frac{dy}{dN}& = y \left[\sqrt{3}\lambda x\sqrt{y}+3\left( 1+s-y\sqrt{1-x^{2}}\right) \right] ,& \\
		\frac{dz}{dN} & =3\left[-s+z\left( s-y\sqrt{1-x^{2}}\right)\right],&
		\\
		\frac{ds}{dN}& =-3 \left[\frac{\left( z+s\right) \left\lbrace  \gamma z+n\left( \gamma z-s\right) \right\rbrace }{z}-s\left( 1+s-y\sqrt{1-x^{2}}\right)\right]  ,& \\
		&~~\label{autonomous_system}
	\end{split}
\end{eqnarray}
Here, we assume $\frac{V'}{V\sqrt{V}}=\lambda$ defines the steepness of the potential function. In this paper, we consider the $\lambda$ to be constant and it becomes a free model parameter. $\lambda$=constant characterises the inverse square potential ($V\propto \frac{1}{\phi^{2}}$) for the tachyon field when $\lambda\neq 0$. On the other hand, $\lambda=0$ corresponds to constant potential which implies that the tachyon leads to cosmological constant like fluid.  $ N=\ln a $ is the e-folding parameter taken to be independent variable. It is clear that the autonomous system (\ref{autonomous_system}) has a mathematical singularity at $z=0$. As a result, the system fails to give any physical information at $z=0$, i.e., at $\Omega_{CG}=0$. In the phase space $x-y-z-s$, we cannot study the properties of solutions lying on $z=0$ plane. In order to remove this singularity, we multiply the right hand sides of the system (\ref{autonomous_system}) by $z$ and in this operation the geometry will be same as before. This operation allows us to analyse
the solutions on $z=0$ plane without changing the qualitative dynamical features of the system in the other regions of the phase space for $z>0$. After applying this technique, we obtain the modified autonomous system as
\begin{eqnarray}
	\begin{split}
		\frac{dx}{dN}& =z\left( x^{2}-1\right) \left( 3x+\sqrt{3}\lambda\sqrt{y}\right)  ,& \\
		\frac{dy}{dN}& = yz \left[\sqrt{3}\lambda x\sqrt{y}+3\left( 1+s-y\sqrt{1-x^{2}}\right) \right] ,& \\
		\frac{dz}{dN} & =3z\left[-s+z\left( s-y\sqrt{1-x^{2}}\right)\right],&
		\\
		\frac{ds}{dN}& =-3 \left[\left( z+s\right) \left\lbrace  \gamma z+n\left( \gamma z-s\right) \right\rbrace-sz\left( 1+s-y\sqrt{1-x^{2}}\right)\right]  ,& \\
		&~~\label{autonomous_system1}
	\end{split}
\end{eqnarray}
Here, the singularity is removed and is now regular at $z=0$ plane.
Cosmological parameters associated to this model are interpreted in terms of the dynamical variables as follows :\\
The density parameters for tachyonic scalar field is defined as  $\Omega_{\phi}=\frac{\rho_{\phi}}{3H^2}$. It, by using Eqn.(\ref{energy density scalar}), will take the form: $\Omega_{\phi}=\frac{V(\phi)}{3H^{2} \sqrt{1-\dot{\phi}^{2}}}$. The density parameter for the modified Chaplygin gas is: $\Omega_{CG}=\frac{\rho_{CG}}{3H^2}$. Now, by using the dynamical variables defined in Eqn.(\ref{dynamical_variables}), the density parameters for tachyonic scalar field and for modified Chaplygin gas will take the following forms:
\begin{equation}\label{density_parameter}
\Omega_{\phi}=\frac{y}{\sqrt{1-x^{2}}},
\end{equation}
and
\begin{equation}\label{density_parameter_f}
\Omega_{CG}=z
\end{equation}
respectively. 

The equation of state parameter for Tachyonic scalar field is
\begin{equation}\label{eqn_of_state_parameter}
\omega_{\phi}=x^{2}-1
\end{equation}
and the equation of state parameter for Chaplygin gas is
\begin{equation}
\omega_{CG}=\frac{s}{z}.
\end{equation}
The effective equation of state or the total equation of state parameter for the model (as defined in Eqn.(\ref{equations})) in terms of dynamical variables reads as
\begin{equation}\label{eff_eqn_of_state_parameter}
\omega_{eff}~ (\equiv \omega_{total})=\frac{p_{total}}{\rho_{total}}= \frac{p_{\phi}+p_{CG}}{\rho_{\phi}+\rho_{CG}}=s-y\sqrt{1-x^{2}},
\end{equation}

and the deceleration parameter for the model takes the form
\begin{equation}\label{dec_parameter}
q=-1+\frac{3}{2}(1+\omega_{eff})=\frac{1}{2}\left\lbrace 1+3\left( s-y\sqrt{1-x^{2}}\right) \right\rbrace 
\end{equation}
Acceleration of the universe is obtained for the condition:~~~$q<0$~~~or~~~ $\omega_{eff}<-\frac{1}{3}$
and the deceleration for:~~~$q>0$~~~{\it i.e.}~~~ $\omega_{eff}>-\frac{1}{3}.$
Friedmann equation (\ref{Friedmann1}) gives the constraint in dynamical variables in the phase space. The restriction for the model is
\begin{equation}\label{constraint}
	z+\frac{y}{\sqrt{1-x^{2}}}=1.
\end{equation}

\section{Phase space analysis of autonomous system (\ref{autonomous_system1}):}
\label{phase space autonomous system}
In this section, we shall extract the critical points from the $4D$ autonomous system (\ref{autonomous_system1}) and we shall investigate the local  stability of the model. For hyperbolic critical points linear stability theory is employed to find their stability. On the other hand, we apply numerical computational method to find their stability. Critical points and their corresponding physical parameters are presented in the table (\ref{physical_parameters}). Eigenvalues of the linearized Jacobian matrix are displayed in the Table \ref{eigenvalues1}.\\

The critical points for the system (\ref{autonomous_system1}) are the following
{\bf 
\begin{itemize}
\item  I.  Critical Point : $P_{1}=(0,~0,~1,~-1)$
\item  II. Critical Point : $ P_{2}=(0,~0,~1,~\gamma)$

\item  III. Critical Point : $P_{3\pm}=(\pm1,~0,~1,~-1)$

\item  IV. Critical Point : $P_{4\pm}=(\pm1,~0,~1,~\gamma)$

\item  V. Critical Point : $P_{5\pm}=(\pm1,~\frac{3(1+\gamma)^{2}}{\lambda^{2}},~1,~\gamma)$

\item  VI. Set of Critical Points : $ P_{6}=(x_c,~y_c,~0,~0)$

\item  VII. Set of Critical Points : $ P_{7}=(x_c,~y_c~,0,~s_c)$

\item  VIII. Set of Critical Points : $ P_{8\pm}=(\pm1,~y_c,~1,~-1)$

\item  IX. Set of Critical Points : $ P_{9}=(0,~1-z_c,~z_c,~-z_c)$

\item  X. Set of Critical Points : $ P_{10\pm}=(\pm1,~\frac{3}{\lambda^{2}},~z_c,~0)$

\item  XI. Set of Critical Points : $ P_{11}=(x_c,~0,~z_c,~0)$
\end{itemize}

}
Here, $\lambda$ and $\gamma$ are free parameters of the model. $x_c,~y_c$ takes any real values in the interval $[-1,~ 1]$ and $z_c$ takes any real values in the interval $[0,~ 1]$.

	\begin{table}[tbp] \centering
		\caption{The Critical Points and the corresponding physical parameters  are presented.}%
		\begin{tabular}
			[c]{ccccccccc}\hline\hline
			\textbf{Critical Points}&$\mathbf{\omega_{\phi}}$& $\mathbf{\Omega_{\phi}}$ & $\mathbf{\omega_{CG}}$ &
			$\mathbf{\Omega_{CG}}$ & $\mathbf{\omega_{eff}}$ & $q$ & Existence &
			\\\hline
			$P_{1}  $ & $-1$ & $0$ &
			$-1$ & $1$ & $-1$ & $-1$  & $\forall \lambda, n, \gamma, \beta$ \\
			$P_{2}  $ & $-1$ & $0$ &
			$\gamma$ & $1$ & $\gamma$ & $\frac{1}{2}(1+3\gamma)$ & $\forall \lambda, n, \gamma, \beta$\\
			$P_{3\pm}  $ & $0$ & Undefined &
			$-1$ & $1$ & $-1$ & $-1$ & $\forall \lambda, n, \gamma, \beta$ \\
			$P_{4\pm}  $ & $0$ & Undefined &
			$\gamma$ & $1$ & $\gamma$ & $\frac{1}{2}(1+3\gamma)$ & $\forall \lambda, n, \gamma, \beta$\\
			$P_{5\pm}  $ & $0$ & $\infty$ &
			$\gamma$ & $1$ & $\gamma$ & $\frac{1}{2}(1+3\gamma)$ & $\forall n, \gamma, \beta, \lambda\neq 0$\\
			$P_{6}  $ & $x_{c}^{2}-1$ & $\frac{y_c}{\sqrt{1-x_{c}^{2}}}$ &
			Undefined & $0$ & $-y_{c} \sqrt{1-x_{c}^{2}}$ & $\frac{1}{2}(1-3y_{c} \sqrt{1-x_{c}^{2}})$ & $\forall \lambda, n, \gamma, \beta$, $x_{c}^{2}+y_{c}^{2}=1$\\
			$P_{7}  $ & $x_{c}^{2}-1$ & $\frac{y_c}{\sqrt{1-x_{c}^{2}}}$ &
			$\infty$ & $0$ & $s_{c}-y_{c} \sqrt{1-x_{c}^{2}}$ & $\frac{1}{2}\left\lbrace 1+3(s_{c}-y_{c} \sqrt{1-x_{c}^{2}})\right\rbrace $ & $\forall \lambda, \gamma, \beta, n=0,s_{c}\neq0, x_{c}^{2}+y_{c}^{2}=1$\\
			$P_{8\pm}  $ & $0$ & $\infty$ &
			$-1$ & $1$ & $-1$ & $-1$ & $\forall n, \gamma, \beta, \lambda=0$\\
			$P_{9}  $ & $-1$ & $1-z_{c}$ &
			$-1$ & $z_{c}$ & $-1$ & $-1$ & $\forall n, \gamma, \beta, \lambda=0, z_{c}\in (0,1]$\\
			$P_{10\pm}  $ & $0$ & $\infty$ &
			$0$ & $z_{c}$ & $0$ & $\frac{1}{2}$ & $\forall n, \beta, \gamma=0, \lambda\neq0, z_{c}\in (0,1]$\\
			$P_{11}  $ & $x_{c}^{2}-1$ & $0$ &
			$0$ & $z_c$ & $0$ & $\frac{1}{2}$ & $\forall \lambda, n, \beta, \gamma=0, z_c=1$
			
			\\\hline\hline
		\end{tabular}
		\label{physical_parameters} \\
		
\end{table}%
%

\begin{table}[tbp] \centering
	\caption{The eigenvalues of the linearized system (\ref{autonomous_system1})}%
	\begin{tabular}
		[c]{ccccccccc}\hline\hline
		\textbf{Critical Points} & $\mathbf{\mu_{1}}$ &
		$\mathbf{\mu_{2}}$ & $\mathbf{\mu_{3}}$ & $\mu_{4}$ & Stability \\\hline
		$P_{1}  $ & $-3$ & $0$ & $-3$ & $-3 (\gamma +1) (n+1)$ &  (See text) \\
		$P_{2}  $ & $-3$ & $3 \gamma$ & $3 (\gamma +1)$ & $3 (\gamma +1) (n+1)$  & Saddle\\
		$P_{3\pm}  $ & $6$ & $-3$ & $0$ & $-3 (\gamma +1) (n+1)$  & Saddle\\
		$P_{4\pm}  $ & $6$ & $3 \gamma$ & $3 (\gamma +1)$ & $3 (\gamma +1) (n+1)$ & Repeller\\
		$P_{5+}  $ & $3 \gamma$ & $3 (\gamma +1) (n+1)$ & $6(2+\gamma)$ & $\frac{15}{2}(1+\gamma)$ & Repeller\\
		$P_{5-}  $ & $3 \gamma$ & $3 (\gamma +1) (n+1)$ & $-6\gamma$ & $-\frac{3}{2}(1+\gamma)$ & Saddle\\
		$P_{6}  $ & $0$ & $0$ & $0$ & $0$ & (See text) \\
		$P_{7}  $ & $0$ & $0$ & $-3s_c$ & $0$ & (See text)\\
		$P_{8\pm}  $ & $6$ & $-3$ & $0$ & $-3 (\gamma +1) (n+1)$  & Saddle\\
		$P_{9}  $ & $-3 z_c$ & $-3 z_c$ & $0$ & $-3 z_c (\gamma +1) (n+1)$ & Stable for $z_{c}\in (0,1]$\\
		$P_{10+}  $ & $0$ & $3z_{c}(1+n)$ & $12z_c$ & $\frac{15}{2}z_c$ & Repeller for $z_{c}\in (0,1]$\\
		$P_{10-}  $ & $0$ & $3z_{c}(1+n)$ & $0$ & $-\frac{3}{2}z_c$ & Saddle for $z_{c}\in (0,1]$ \\
		$P_{11}  $ & $0$ & $3$ & $3(3x_{c}^{2}-1)$ & $3(1+n)$ & Repeller for $\left( -1\leq x_{c}<-\frac{1}{\sqrt{3}}~\mbox{or}~ \frac{1}{\sqrt{3}}<x_{c}\leq 1\right) $ \\ &  & & & & and saddle otherwise
		
		\\\hline\hline
	\end{tabular}
	\label{eigenvalues1}
\end{table}%
\subsection{LOCAL STABILITY CRITERIA OF CRITICAL POINTS :}\label{local_stability}
Local stability of the critical points can be found by perturbing the system up-to first order about the critical points ({\it i.e.,} by linearizing the system around the points). To find the local stability, we need to find out the eigenvalues of perturbed matrix at the critical points. Stability criteria for the points of the four dimensional autonomous system (\ref{autonomous_system1}) are carried out by numerical computational method which are presented below in this subsection. 

\begin{itemize}
\item The critical point $P_{1}$ corresponds to an accelerated Chaplygin gas dominated solution where the Chaplygin fluid behaves as cosmological constant ($\Omega_{CG}=1,~\omega_{CG}=-1$) like fluid. And it evolves in cosmological constant era ($\omega_{eff}=-1$) as if it can mimic the de Sitter like solution in the phase space. From the table \ref{eigenvalues1}, we observe that the point $P_{1}$ having exactly one vanishing eigenvalue, is a non-hyperbolic type of critical point and the stability of this type of point can be determined by employing the centre manifold theory  \cite{Wiggins}. However, in the present paper, we shall simply refer to the numerical computational method by using perturbating the trajectories 
to determine the stability of the point as $N\longrightarrow \infty$. This technique has been found to be quite successful in literature for acquiring the stability behavior of critical points of the system \cite{Dutta,Zonu,Dutta1,Dutta2,Dutta3,Dutta4}. Now, we perturb the trajectories near the coordinates of the critical point and sketch the projections of the phase trajectories along the axes $x$, $y$, $z$ and $s$. The projection of phase trajectories are shown in fig. (\ref{P1-perturbation}) where we choose the values of initial conditions that are very near to the co-ordinates of the critical point. It is observed that the sub figures \ref{fig:Stable_Nx_P1}, \ref{fig:Stable_Nz_P1} and \ref{fig:Stable_Ns_P1} in fig. (\ref{P1-perturbation}) exhibit the phase-space trajectories approach $x = 0$, $z=1$ and $s=-1$ respectively as $N \rightarrow \infty$. However, in sub figure \ref{fig:Stable_Ny_P1}, it can be seen that trajectories do not approach the coordinate axis $y=0$. Hence, from the above, it is concluded that behavior of the perturbed system near the critical point $P_{1}$ indicates that the point is unstable in the phase space $x-y-z-s$. Therefore, the point describes the transient de Sitter like solution. It is to be noted that the density of the MCG for this critical point will approach to a constant value. As Eqn.(\ref{MCG EoS 1}) gives: $\omega_{CG}=\frac{p_{CG}}{\rho_{CG}}=\gamma-\frac{\beta}{\rho^{n+1}_{CG}}$. Hence, $\omega_{CG}=-1$ will lead to $\rho_{CG}=\left(\frac{\beta}{1+\gamma} \right)^{\frac{1}{n+1}}$. Thus, the behaviour of the chaplygin gas approaches that of LIVE with a constant
density which is represented by cosmological constant. Furthermore, the condition $\Omega_{CG}=1$ corresponds to $\frac{\rho_{CG}}{3H^2}=1$ giving $H=\frac{1}{\sqrt{3}} \left(\frac{\beta}{1+\gamma} \right)^{\frac{1}{2(n+1)}}$. 

\begin{figure}
	\centering
	\subfigure[]{%
		\includegraphics[width=3.9cm,height=5.5cm]{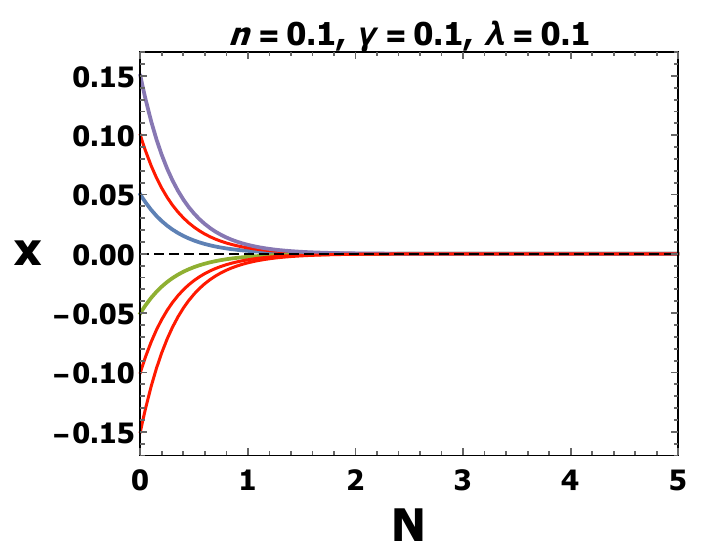}\label{fig:Stable_Nx_P1}}
	\qquad
	\subfigure[]{%
		\includegraphics[width=3.9cm,height=5.5cm]{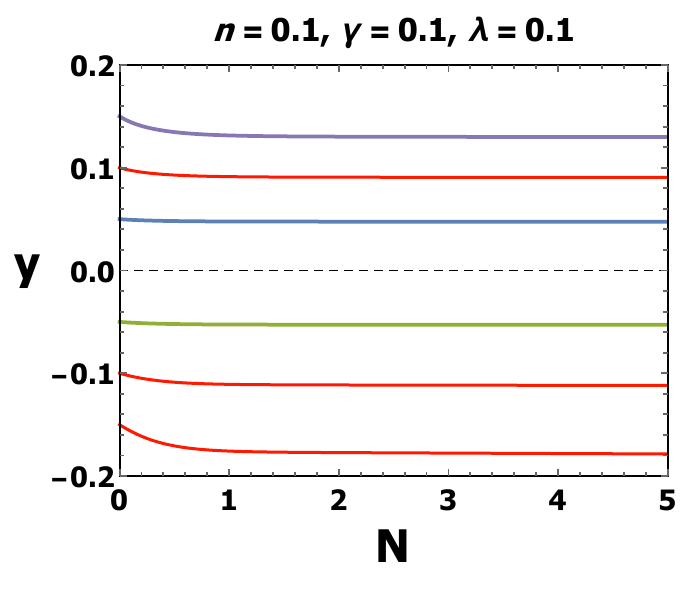}\label{fig:Stable_Ny_P1}}
	\qquad
	\subfigure[]{%
		\includegraphics[width=3.9cm,height=5.5cm]{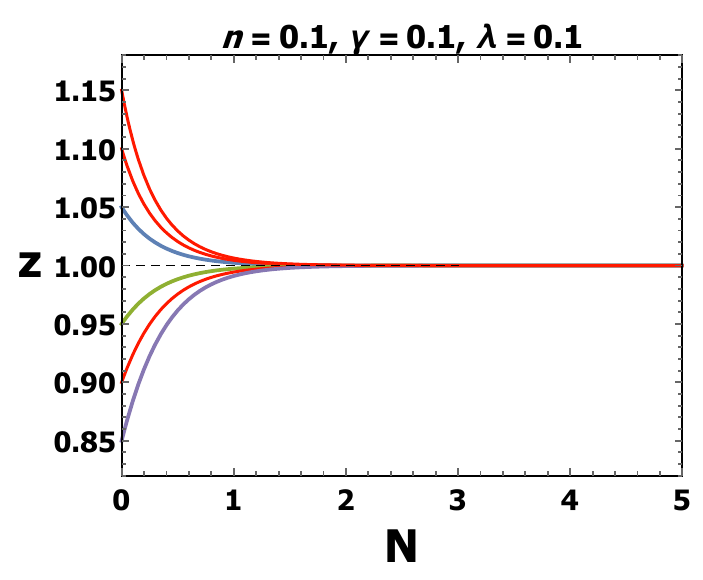}\label{fig:Stable_Nz_P1}}
	\qquad
	\subfigure[]{%
		\includegraphics[width=3.9cm,height=5.5cm]{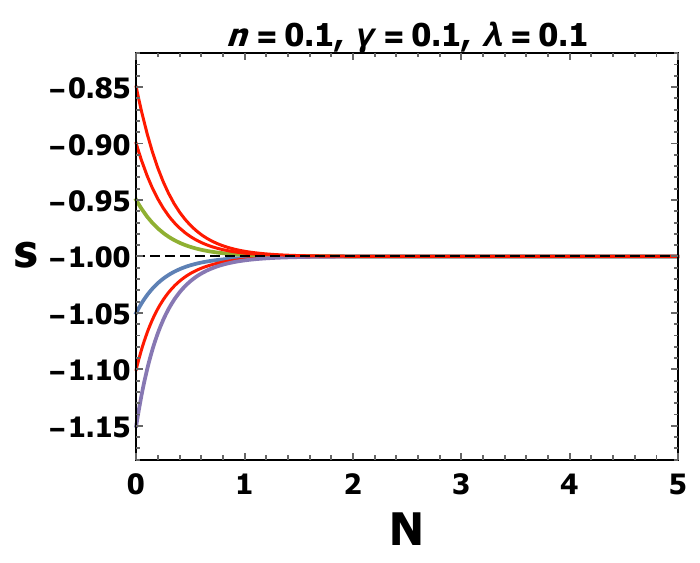}\label{fig:Stable_Ns_P1}}
	\caption{The figure shows the projection of time evolution of the phase space trajectories of perturbation of the autonomous system (\ref{autonomous_system1}) along the x,y,z and s axes for the critical point $P_{1}$ for parameter values $n=0.1$, $\gamma=0.1$ and $\lambda=0.1$ which
		determine the unstability of $P_{1}$.}
	\label{P1-perturbation}
\end{figure}


\item The critical point $P_{2}$ corresponds to a decelerated Chaplygin gas dominated solution ($\Omega_{CG}=1,~\omega_{eff}=\gamma$) where the chaplygin fluid behaves as perfect fluid in nature ($\omega_{CG}=\gamma$). It should be mentioned that $\omega_{CG}=\gamma$ requires $\beta=0$ in Eqn.(\ref{MCG EoS 1}).
It is hyperbolic in nature since all the eigenvalues have nonzero real parts and it exhibits the saddle like solution in the phase space (see in Table \ref{eigenvalues1}). For $\gamma=\frac{1}{3}$, the MCG mimics as radiation and the point evolves in radiation dominated era $\omega_{eff}=\frac{1}{3}$ and it has the transient nature. On the other hand, for $\gamma=0$, the MCG behaves as dust fluid ($\omega_{CG}=0$) and the universe evolves in dust era $\omega_{eff}=0$. However, the point is non-hyperbolic type in this case. But, it shows saddle like nature in the phase space. So, it represents the dust dominated intermediate phase of the universe. 

\item  The critical points $P_{3\pm}$ correspond to an accelerated Chaplygin gas dominated solutions where Chaplygin gas behaves as cosmological constant (Since [$\Omega_{CG}=1$, $\omega_{CG}=\omega_{total}=-1$] see Table \ref{physical_parameters}).
However, applying the coordinates of this point in the Eq. (\ref{density_parameter}), the density parameter for the tachyon field ($\Omega_{\phi}$) will be of indeterminant form. As a result, depending on the direction of the trajectories approaching them, the critical points $P_{3\pm}$ can describe tachyon field dominated or MCG dominated solutions in the phase space. When trajectories are very near to the boundary of $x^{2}+y^{2}=1$, they can represent Tachyonic field dominated solutions. Otherwise, when trajectories close to the axis $y = 0$, they can represent the Chaplygin gas dominated solutions. This type of solution can be found in the study of conformally coupled tachyonic dark energy in Ref. \cite{Teixeira 2019}. It is non-hyperbolic in nature since one of eigenvalues is zero (see in Table \ref{eigenvalues1}). Both the points are saddle like solutions in the phase space. Therefore, the points can describe the accelerated transient de Sitter phase of the universe dominated by chaplygin gas with cosmological constant equation of state. It is worthy to mention that the behaviour of the chaplygin gas for the critical points approaches that of LIVE with a constant density which is represented by cosmological constant. Since the Eqn.(\ref{MCG EoS 1}) gives: $\omega_{CG}=\frac{p_{CG}}{\rho_{CG}}=\gamma-\frac{\beta}{\rho^{n+1}_{CG}}$ and $\omega_{CG}=-1$ will lead to $\rho_{CG}=\left(\frac{\beta}{1+\gamma} \right)^{\frac{1}{n+1}}$.  Furthermore, the condition $\Omega_{CG}=1$ corresponds to $\frac{\rho_{CG}}{3H^2}=1$ giving $H=\frac{1}{\sqrt{3}} \left(\frac{\beta}{1+\gamma} \right)^{\frac{1}{2(n+1)}}$.

\item The critical points $P_{4\pm}$ correspond to an decelerated Chaplygin gas dominated solution. However, similarly to the points $P_{3\pm}$, due to indetermination of Eq. (\ref{density_parameter}), the points can behave tachyon dominated for trajectories closed to $x^{2}+y^{2}=1$ or they can represent chaplygin fluid dominated for $y = 0$. They are hyperbolic in nature. Since all of the eigenvalues have positive real parts, they are always repellers in the phase space $x-y-z-s$. Equation of state parameter for the modified chaplygin gas follows the perfect fluid equation of state: $\omega_{CG}=\gamma$ which requires $\beta=0$ in Eqn.(\ref{MCG EoS 1}).
For $\gamma=\frac{1}{3}$, the points evolve in decelerated radiation era where the points dominated by chaplygin gas which behaves as radiation ($\omega_{CG}=\frac{1}{3},~ \omega_{eff}=\frac{1}{3}$), these are past attractors and represent early evolution of the universe. On the other hand, for $\gamma=0$, the points will evolve in dust dominated decelerated phase when chaplygin gas behaves as dust.

\item The critical points $P_{5\pm}$ correspond to an decelerated Chaplygin gas dominated solution where the equation of state parameter for the modified chaplygin gas takes the perfect fluid equation of state: $\omega_{CG}=\gamma$ which requires $\beta=0$ in Eqn.(\ref{MCG EoS 1}). They are hyperbolic in nature having non vanishing eigenvalues. Since all of the eigenvalues are real valued and positive for $P_{5+}$ it is always repeller. But for critical point $P_{5-}$ some eigen values are positive and some are negative. So, it is a saddle point and $P_{5-}$ is relevant for $\gamma=0$ when the point represents dust dominated (here chaplygin gas mimics as dust $\omega_{CG}=0$) decelerated phase ($q=\frac{1}{2}$) in dust era ($\omega_{eff}=0$) and since the point is saddle solution, it represents the intermediate evolution of the universe. On the other hand, for $\gamma=\frac{1}{3}$, the point $P_{5+}$ is relevant in early evolution when it represents the radiation dominated source which evolves in radiation era ($\omega_{CG}=\frac{1}{3},~\omega_{eff}=\frac{1}{3}$).

\item The set of critical points $P_{6}$ corresponds to Tachyonic field dominated solution ($\Omega_{\phi}=1,~\Omega_{CG}=0$) due to its existence condition i.e., $x_{c}^2 +y_{c}^2 =1$. Tachyon fluid behaves as perfect fluid in this case (since $\omega_{\phi}=x_{c}^2 -1$). It behaves as quintessence for $0<x_{c}^2 <\frac{2}{3}$ and as cosmological constant for $x_{c}=0$, but, the fluid is unable to give phantom nature. Acceleration is possible for this solution is $y_{c} \sqrt{1-x_{c}^2}>\frac{1}{3}$. The point $P_{6}$ is non-hyperbolic in nature since all the eigenvalues are zero. The dynamics near the set is very complicated. We therefore, investigate the stability by using numerical computation Fig. The point $P_{6}$ shows the stability of the set $P_6$. It is exhibited from the sub fig. \ref{fig:Stable_Nz_P6} that the phase space trajectories approach $z = 0$ and in sub-fig. \ref{fig:Stable_Nx_P6}, \ref{fig:Stable_Ny_P6}
the trajectories started from any values of $x$ and $y$ remains almost constant as $N \rightarrow \infty$. On the other hand, the sub-fig. \ref{fig:Stable_Ns_P6} shows that the trajectories do not approach to $s=0$ (the co-ordinate of set $P_6$). Therefore, we can conclude by observing the behavior of the perturbed system near the set $P_{6}$ that the set is unstable in phase space. Therefore, the tachyon field dominated solution cannot provide the late time evolution.

\begin{figure}
	\centering
	\subfigure[]{%
		\includegraphics[width=3.9cm,height=5.5cm]{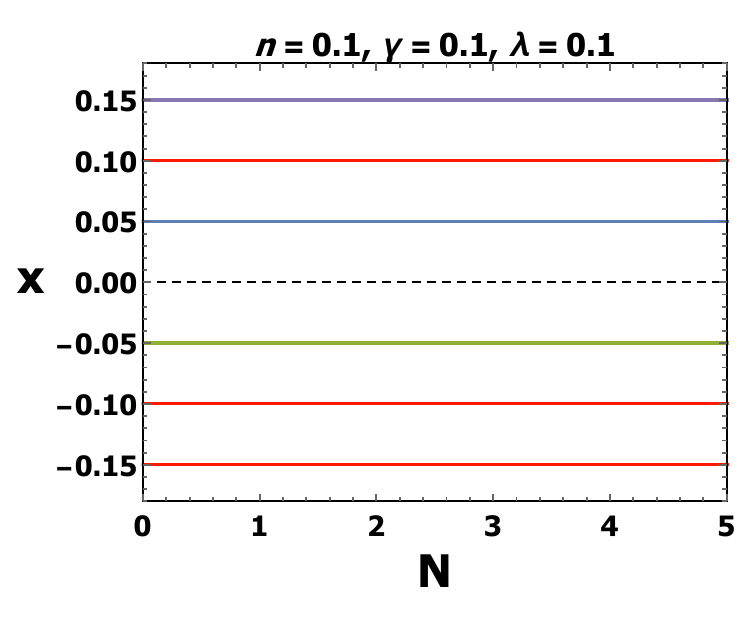}\label{fig:Stable_Nx_P6}}
	\qquad
	\subfigure[]{%
		\includegraphics[width=3.9cm,height=5.5cm]{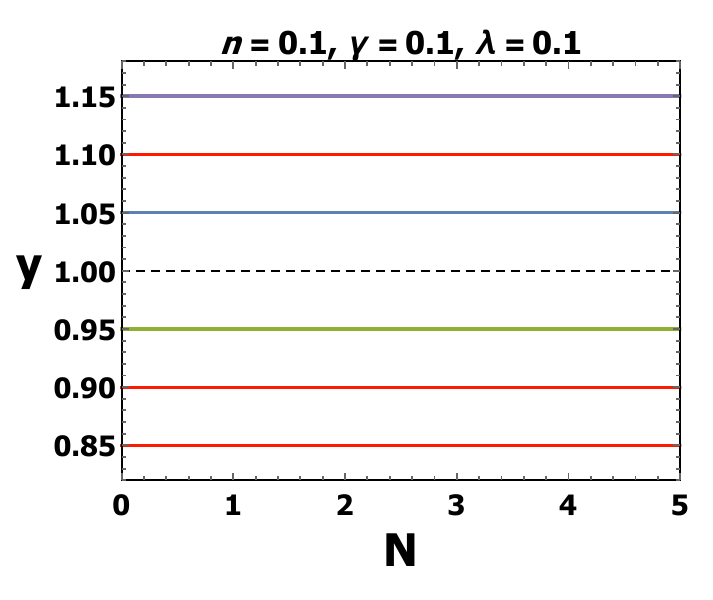}\label{fig:Stable_Ny_P6}}
	\qquad
	\subfigure[]{%
		\includegraphics[width=3.9cm,height=5.5cm]{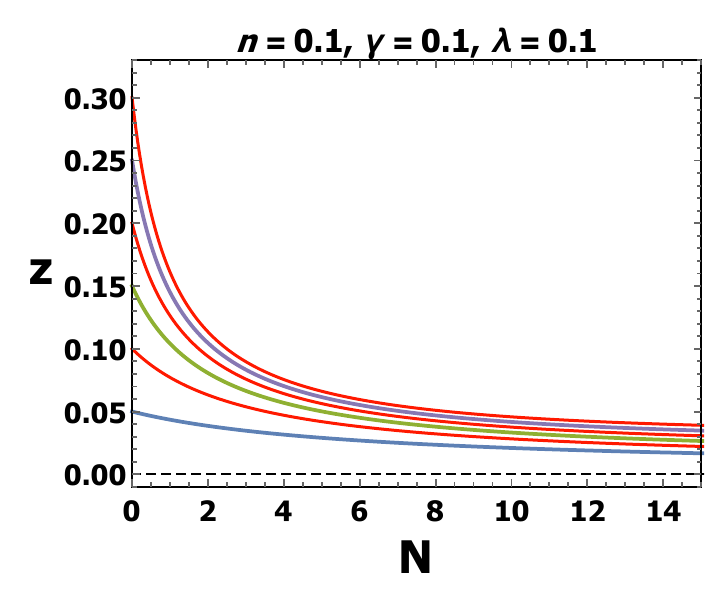}\label{fig:Stable_Nz_P6}}
	\qquad
	\subfigure[]{%
		\includegraphics[width=3.9cm,height=5.5cm]{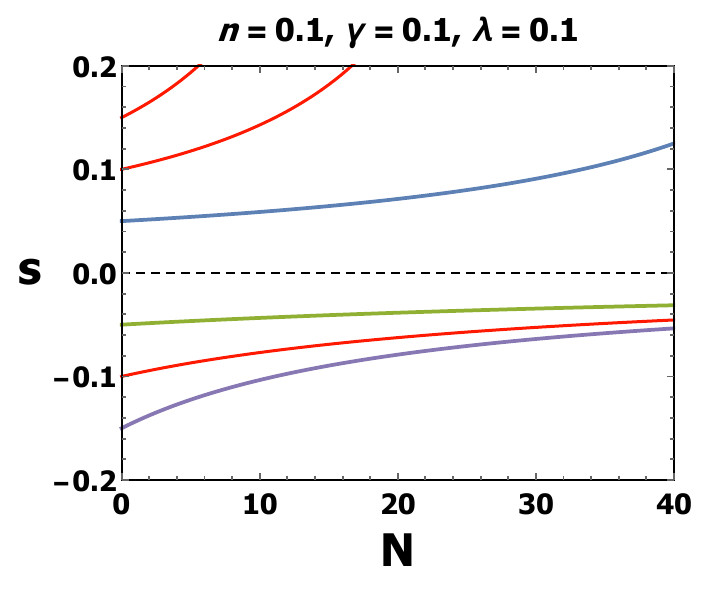}\label{fig:Stable_Ns_P6}}
	\caption{The figure shows the projection of time evolution of the phase space trajectories of perturbation of the autonomous system (\ref{autonomous_system1}) along the x,y,z and s axes for the critical point $P_{6}$ for parameter values $n=0.1$, $\gamma=0.1$ and $\lambda=0.1$ which 
		determine the unstability of $P_{6}$.}
	\label{P6-perturbation}
\end{figure}

\item Another set of critical points $P_{7}$ corresponds to a Tachyonic field dominated solution and tachyon field behaves similarly as of the set $P_6$. The set $P_7$ is also a non-hyperbolic set of points having three vanishing eigenvalues, so, it is not normally hyperbolic set. Numerical investigation can provide the mechanism for stability of the set. For $s_{c}\neq 0$, it has one dimensional non-empty stable sub manifold when $s_{c}>0$. Fig. \ref{P7-perturbation} shows the stability of the set of points $P_{7}$. 
It is clear from sub-fig. \ref{fig:Stable_Nz_P7} 
that the phase space trajectories approach $z = 0$ and in the sub-figs. \ref{fig:Stable_Nx_P7}, \ref{fig:Stable_Ny_P7} and \ref{fig:Stable_Ns_P7} the trajectories starting from any values of $x$ and $y$ and $s$ remain almost constant as $N \rightarrow \infty$. From the behavior of the perturbated trajectories, it can be concluded that the set of points $P_{7}$ is stable in the phase space. Therefore, the set shows the late-time behaviour of the evolution of the universe. The universe is accelerated near the set for $y_{c} \sqrt{1-x_{c}^2}> s_{c}+\frac{1}{3}$. The set $P_7$ represents the late-time accelerated evolution attracted only in quintessence era ($-1<\omega_{eff}<-\frac{1}{3}$) for the following conditions: $-\sqrt{\frac{2}{3}}<x_c<\sqrt{\frac{2}{3}},~y_{c}=\sqrt{1-x_{c}^{2}},~ 0<s_{c}<-\frac{1}{3}+y_{c} \sqrt{1-x_{c}^{2}}$ and $n=0$. Here, the tachyon field behaves as quintessence like fluid.


\begin{figure}
	\centering
	\subfigure[]{%
		\includegraphics[width=3.9cm,height=5.5cm]{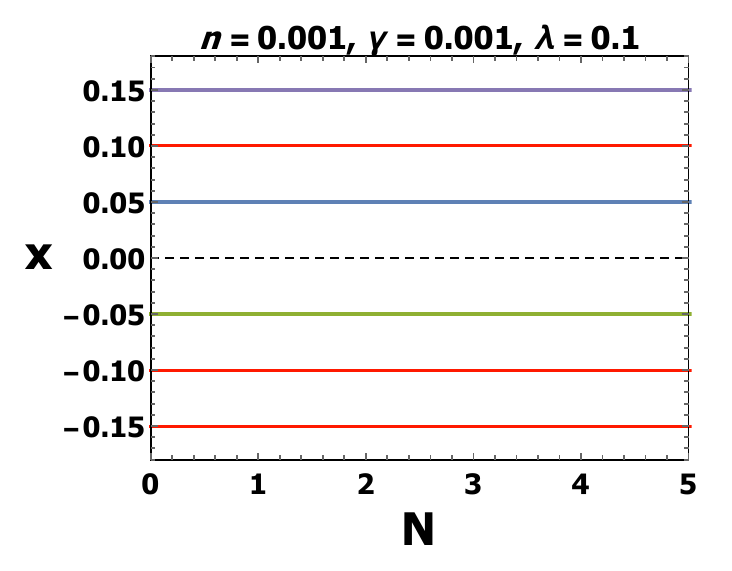}\label{fig:Stable_Nx_P7}}
	\qquad
	\subfigure[]{%
		\includegraphics[width=3.9cm,height=5.5cm]{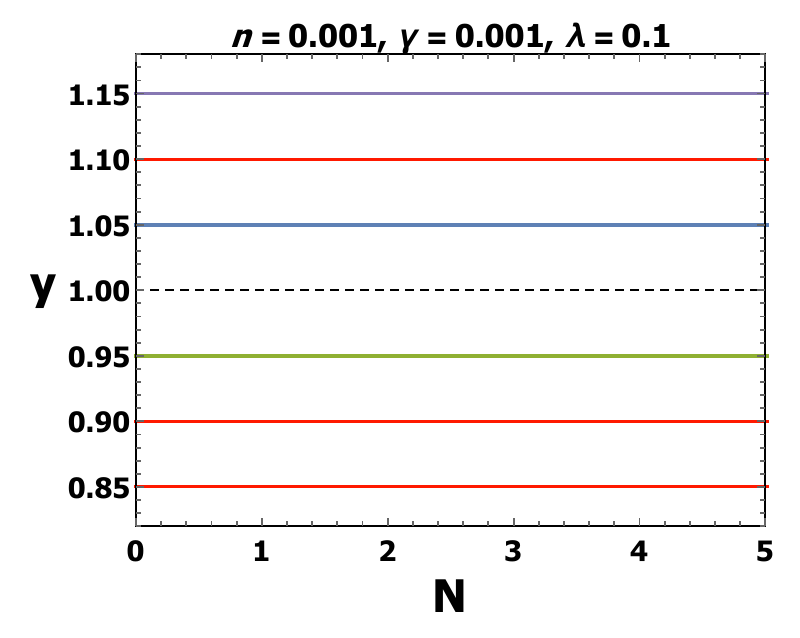}\label{fig:Stable_Ny_P7}}
	\qquad
	\subfigure[]{%
		\includegraphics[width=3.9cm,height=5.5cm]{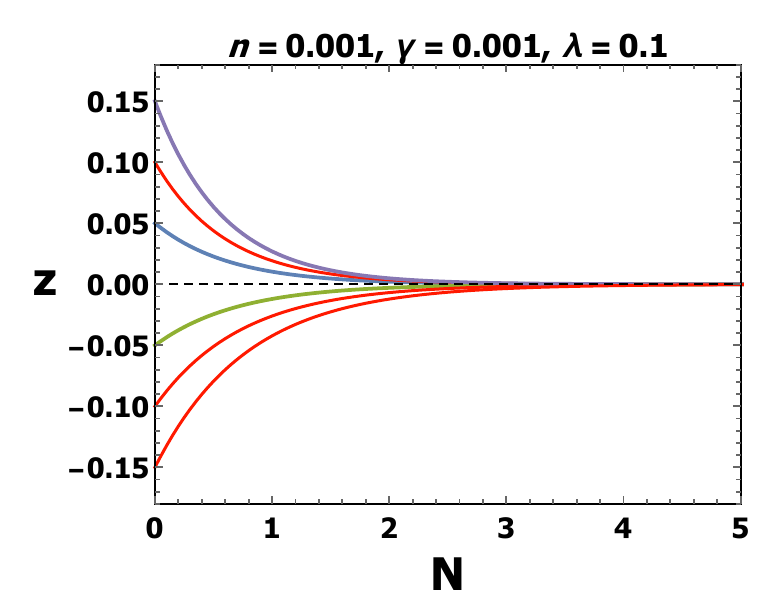}\label{fig:Stable_Nz_P7}}
	\qquad
	\subfigure[]{%
		\includegraphics[width=3.9cm,height=5.5cm]{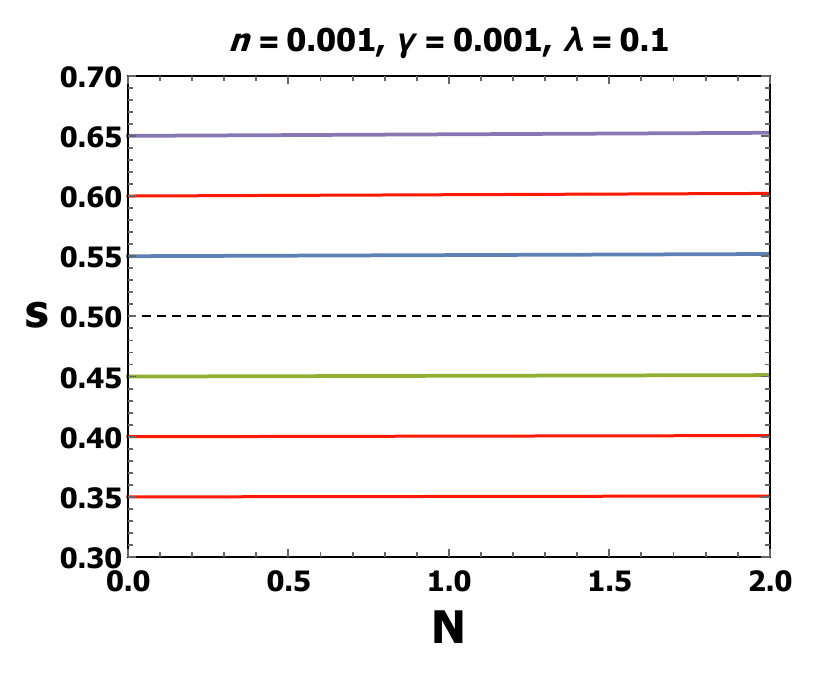}\label{fig:Stable_Ns_P7}}
	\caption{The figure shows the projection of time evolution of the phase space trajectories of perturbation of the autonomous system (\ref{autonomous_system1}) along the x,y,z and s axes for the critical point $P_{7}$ for parameter values $n=0.001$, $\gamma=0.001$ and $\lambda=0.1$ which
		determine the stability of $P_{7}$.}
	\label{P7-perturbation}
\end{figure}

\begin{figure}
	\centering
	\subfigure[]{%
		\includegraphics[width=5.4cm,height=5.4cm]{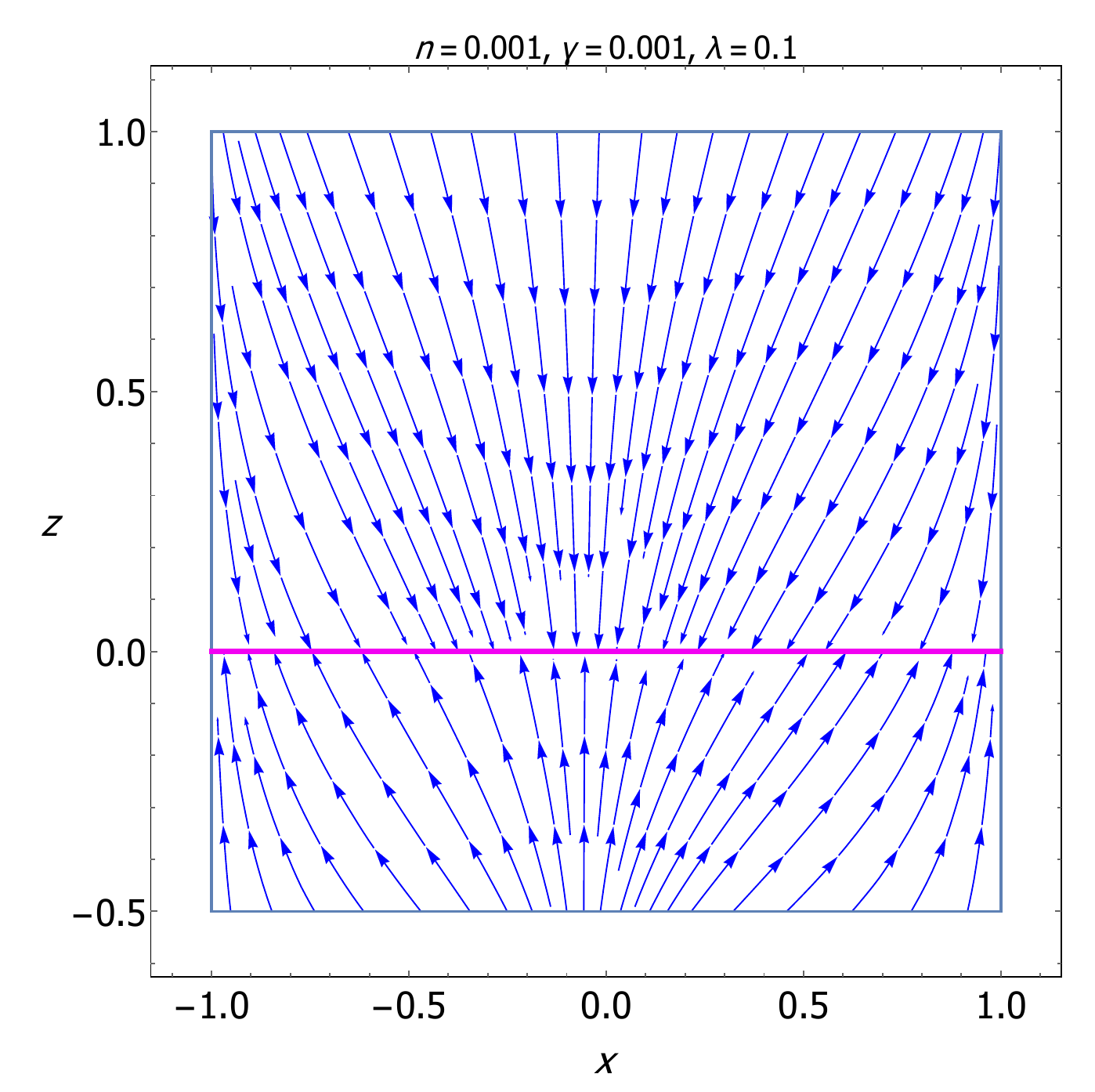}\label{P7_xz}}
	\qquad
	\subfigure[]{%
		\includegraphics[width=5.4cm,height=5.4cm]{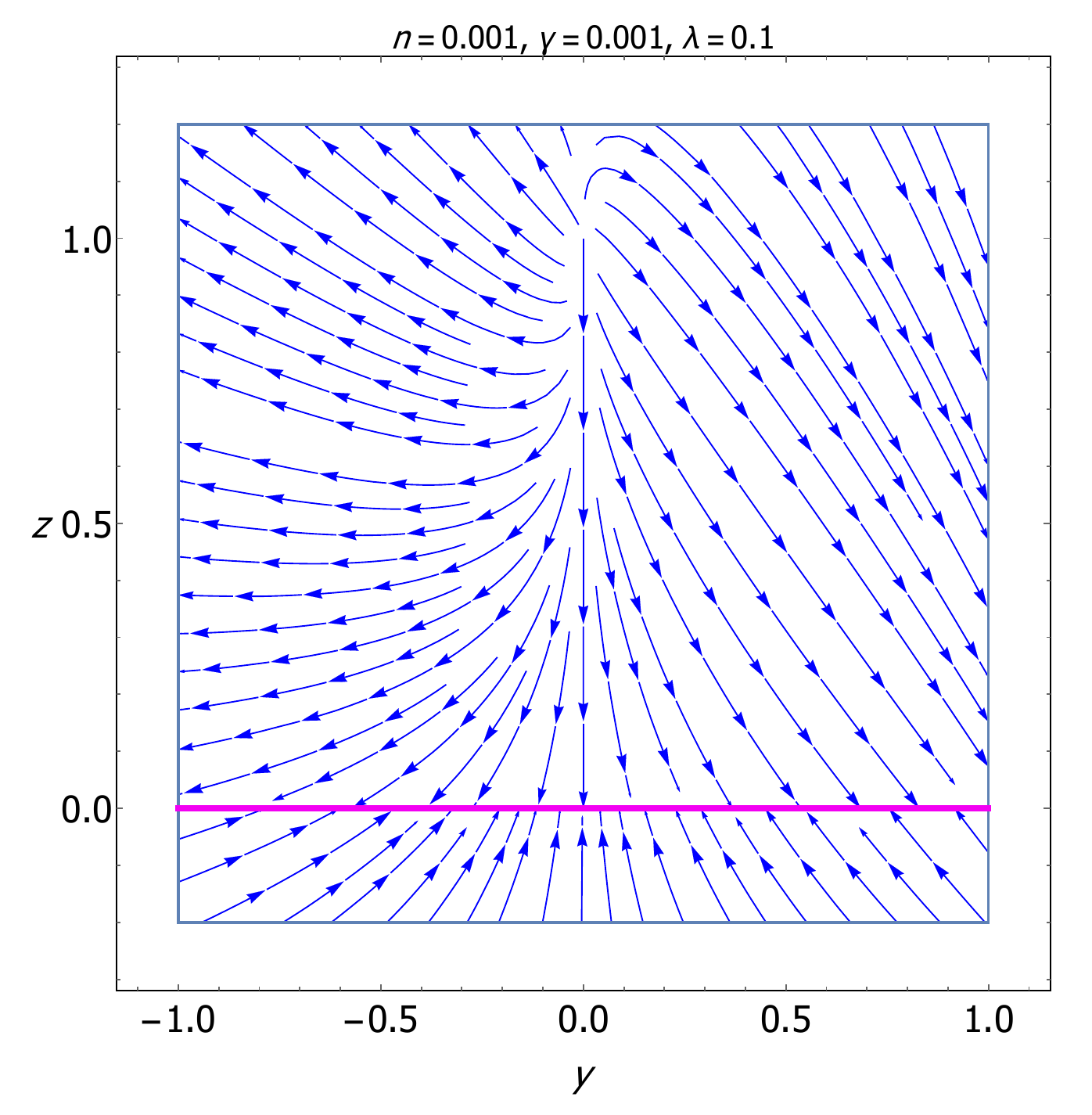}\label{P7_yz}}
	\qquad
	\subfigure[]{%
		\includegraphics[width=5.4cm,height=5.4cm]{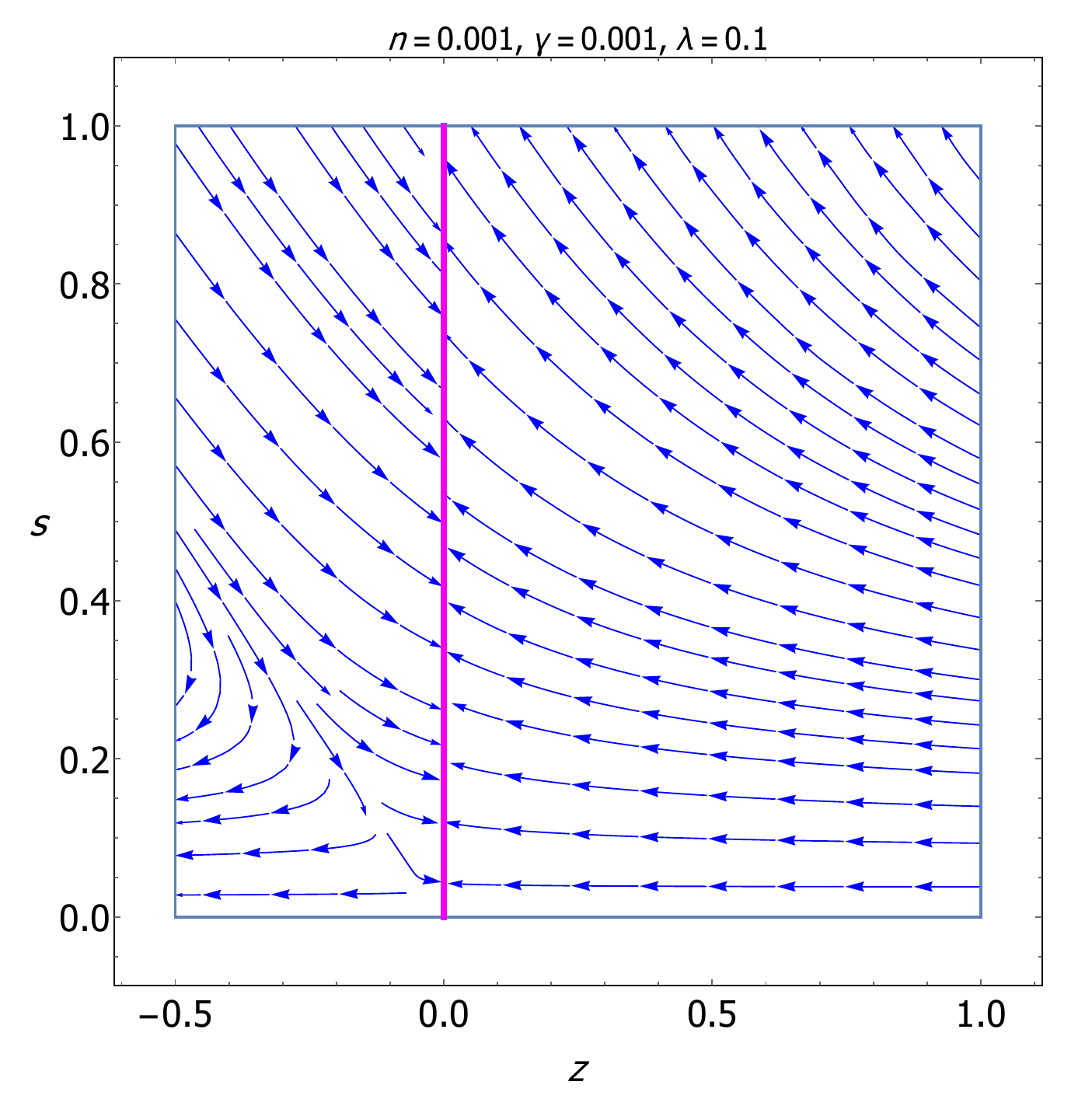}\label{P7_zs}}
	
	\caption{The figure shows the phase space projection of the autonomous system (\ref{autonomous_system1}) on the x-z, y-z and z-s plane in panel \ref{P7_xz}, \ref{P7_yz} and \ref{P7_zs} respectively for the parameter values $n=0.001$, $\gamma=0.001$ and $\lambda=0.1$. In all the panels the set of critical points $P_7$ is shown in the magenta colored line which is attractor in the phase plane.}
	\label{P7_stream}
\end{figure}

\item The sets of critical points $P_{8\pm}$ correspond to an accelerated Chaplygin gas dominated solution where the chaplygin fluid behaves as cosmological constant [$\Omega_{CG}=1$, $\omega_{CG}=\omega_{total}=-1$]. It is interesting to note that the behaviour of the chaplygin gas for the critical points approaches that of LIVE with a constant density which is represented by cosmological constant. Since the Eqn.(\ref{MCG EoS 1}) gives: $\omega_{CG}=\frac{p_{CG}}{\rho_{CG}}=\gamma-\frac{\beta}{\rho^{n+1}_{CG}}$ and $\omega_{CG}=-1$ will lead to $\rho_{CG}=\left(\frac{\beta}{1+\gamma} \right)^{\frac{1}{n+1}}$.  Furthermore, the condition $\Omega_{CG}=1$ corresponds to $\frac{\rho_{CG}}{3H^2}=1$ giving $H=\frac{1}{\sqrt{3}} \left(\frac{\beta}{1+\gamma} \right)^{\frac{1}{2(n+1)}}$.

They are non-hyperbolic in nature sine the sets have one vanishing eigenvalues. Also, the sets have both positive and negative eigenvalues. Therefore, the sets are saddle like solutions in the phase space. The sets are accelerated de Sitter like solutions but they have transient nature in the phase space. So, the sets are not interested in late-times. 

\item The set of critical points $P_{9}$ corresponds to an accelerated scaling solution which may be chaplygin fluid dominated or tachyonic fluid dominated depending on the choice of $z_c$. The set is non-hyperbolic in nature. Moreover, it is normally hyperbolic as it has only one vanishing eigenvalue and it is stable always for $z_{c}\in (0,1] $. A numerical investigation is exhibited in the fig. (\ref{P9-perturbation}) where the sub-fig.\ref{fig:Stable_Nx_P9} shows the perturbation trajectories approach to the axis $x=0$ ($x$-co-ordinate of the set of points $P_9$) and in the sub-figs. \ref{fig:Stable_Ny_P9}, \ref{fig:Stable_Nz_P9} and \ref{fig:Stable_Ns_P9} the perturbed system show that the trajectories starting from different points of $y$, $z$, and $s$ will remain constant which prove that the set of points $P_9$ is stable in the phase space. Interestingly the set can solve the coincidence problem and can describe the late-time accelerated evolution of the universe either dominated by chaplygin fluid or dominated by tachyonic fluid depending on choice of $z_c$. For both the cases, both the fluids behave as cosmological constant. Late time attractor in this case will always be attracted in cosmological constant era. Note that the behaviour of the chaplygin gas for the critical point approaches that of LIVE with a constant density which is represented by cosmological constant. By Eqn.(\ref{MCG EoS 1}) $\omega_{CG}=\frac{p_{CG}}{\rho_{CG}}=\gamma-\frac{\beta}{\rho^{n+1}_{CG}}$ and $\omega_{CG}=-1$ one can obtain $\rho_{CG}=\left(\frac{\beta}{1+\gamma} \right)^{\frac{1}{n+1}}$. For $z_c =1$, the point will completely be dominated by chaplygin gas, i.e., $\Omega_{CG}=1$ which corresponds to $\frac{\rho_{CG}}{3H^2}=1$ giving $H=\frac{1}{\sqrt{3}} \left(\frac{\beta}{1+\gamma} \right)^{\frac{1}{2(n+1)}}$

\begin{figure}
	\centering
	\subfigure[]{%
		\includegraphics[width=3.9cm,height=5.5cm]{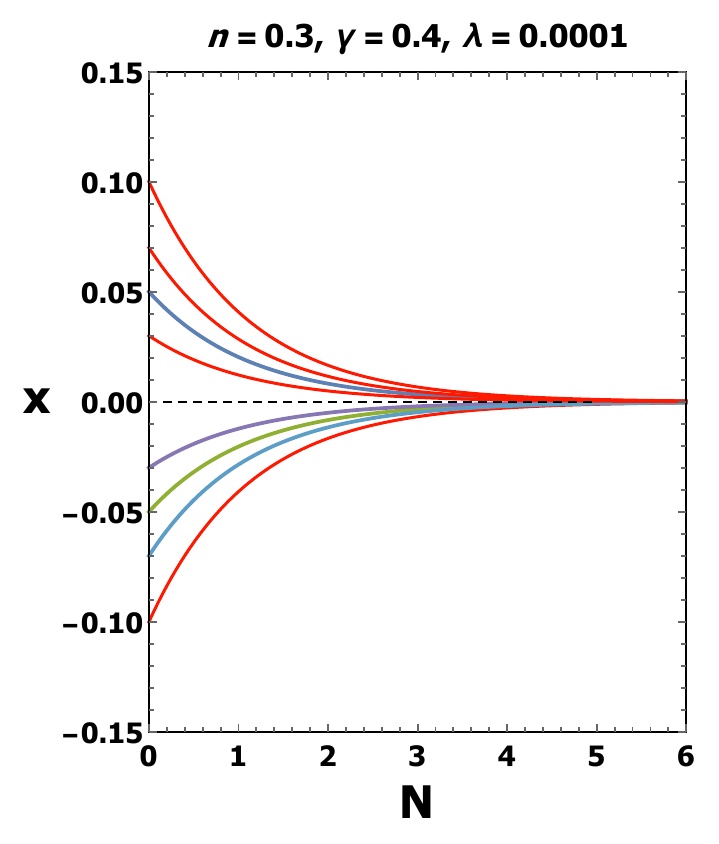}\label{fig:Stable_Nx_P9}}
	\qquad
	\subfigure[]{%
		\includegraphics[width=3.9cm,height=5.5cm]{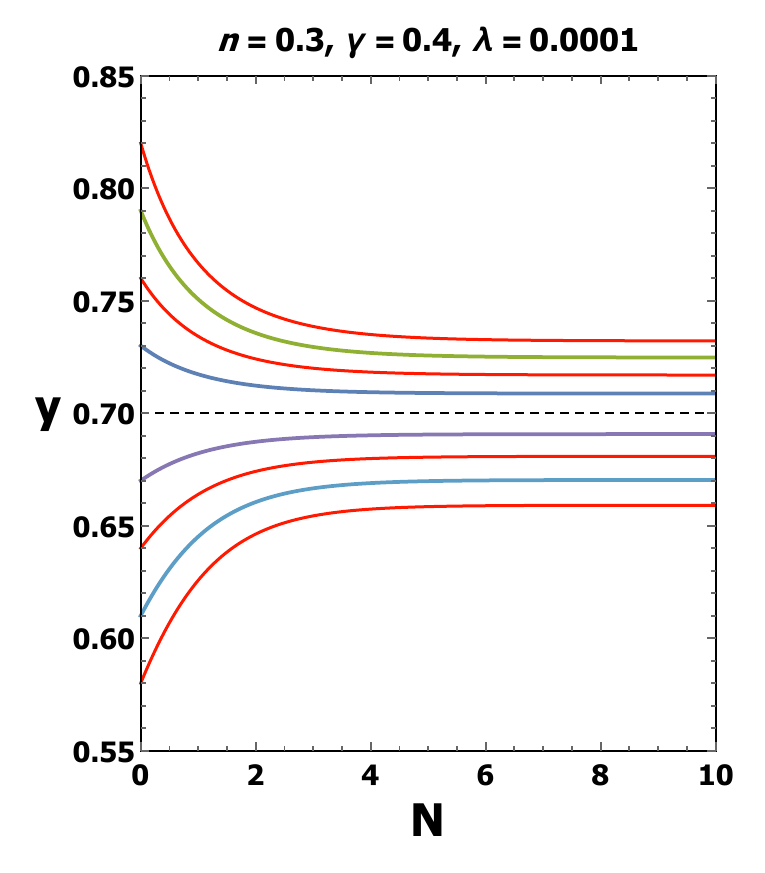}\label{fig:Stable_Ny_P9}}
	\qquad
	\subfigure[]{%
		\includegraphics[width=3.9cm,height=5.5cm]{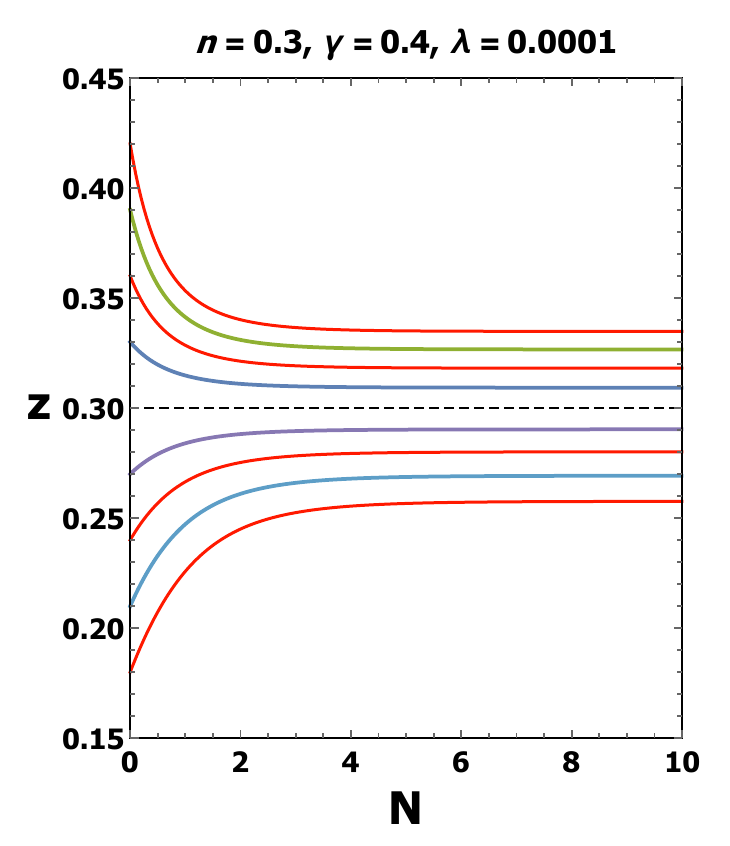}\label{fig:Stable_Nz_P9}}
	\qquad
	\subfigure[]{%
		\includegraphics[width=3.9cm,height=5.5cm]{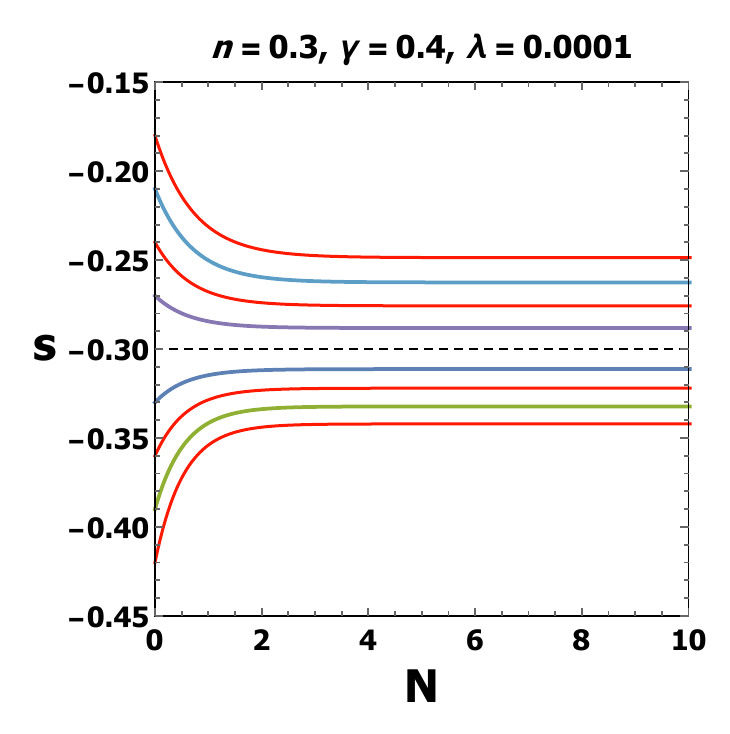}\label{fig:Stable_Ns_P9}}
	\caption{The figure shows the projection of time evolution of the phase space trajectories of perturbation of the autonomous system (\ref{autonomous_system1}) along the x,y,z and s axes for the critical point $P_{9}$ for parameter values $n=0.3$, $\gamma=0.4$ and $\lambda=0.0001$ which
		determine the stability of $P_{9}$.}
	\label{P9-perturbation}
\end{figure}


\begin{figure}
	\centering
	\subfigure[]{%
		\includegraphics[width=5.4cm,height=5.4cm]{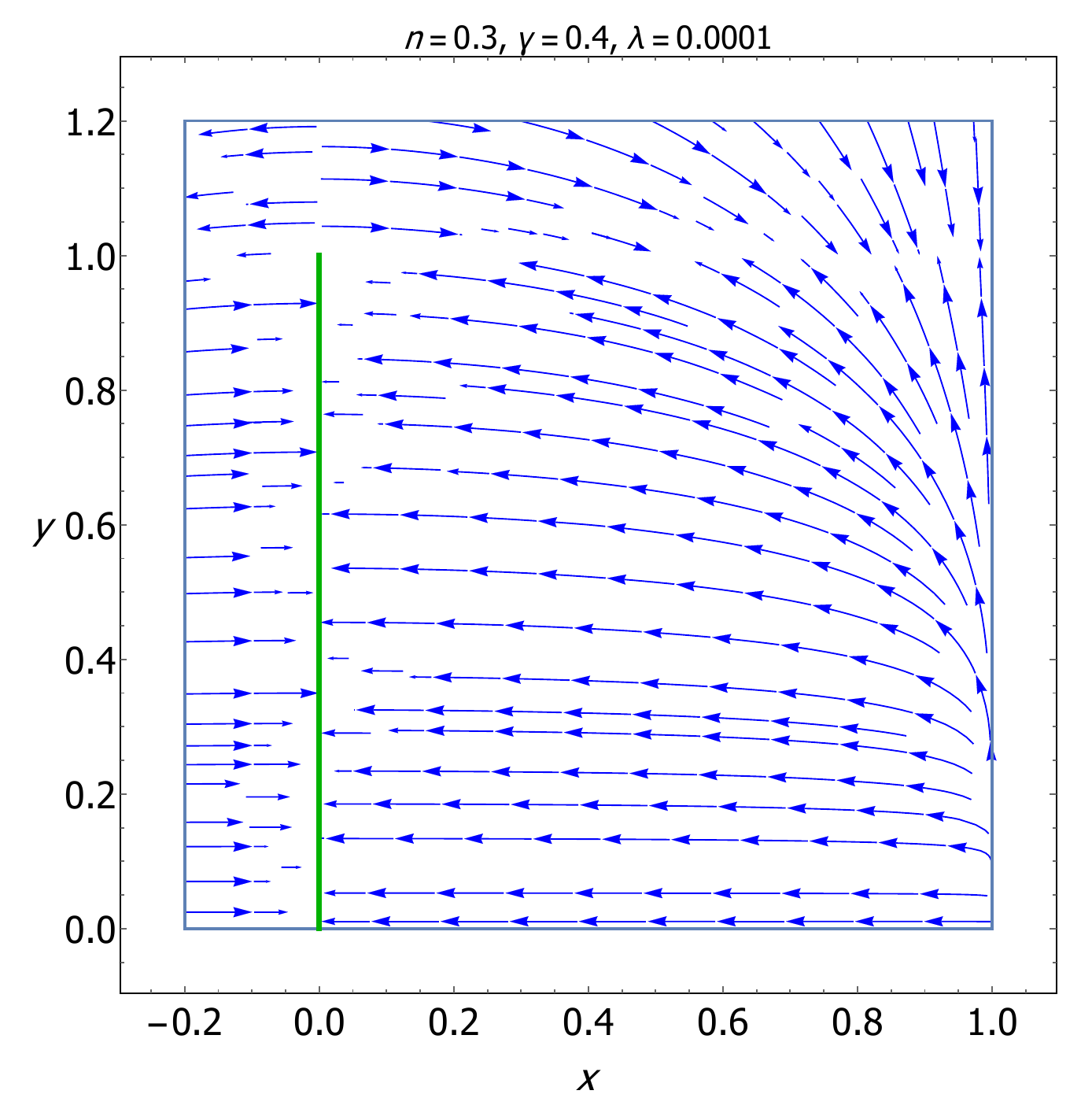}\label{P9_xy}}
	\qquad
	\subfigure[]{%
		\includegraphics[width=5.4cm,height=5.4cm]{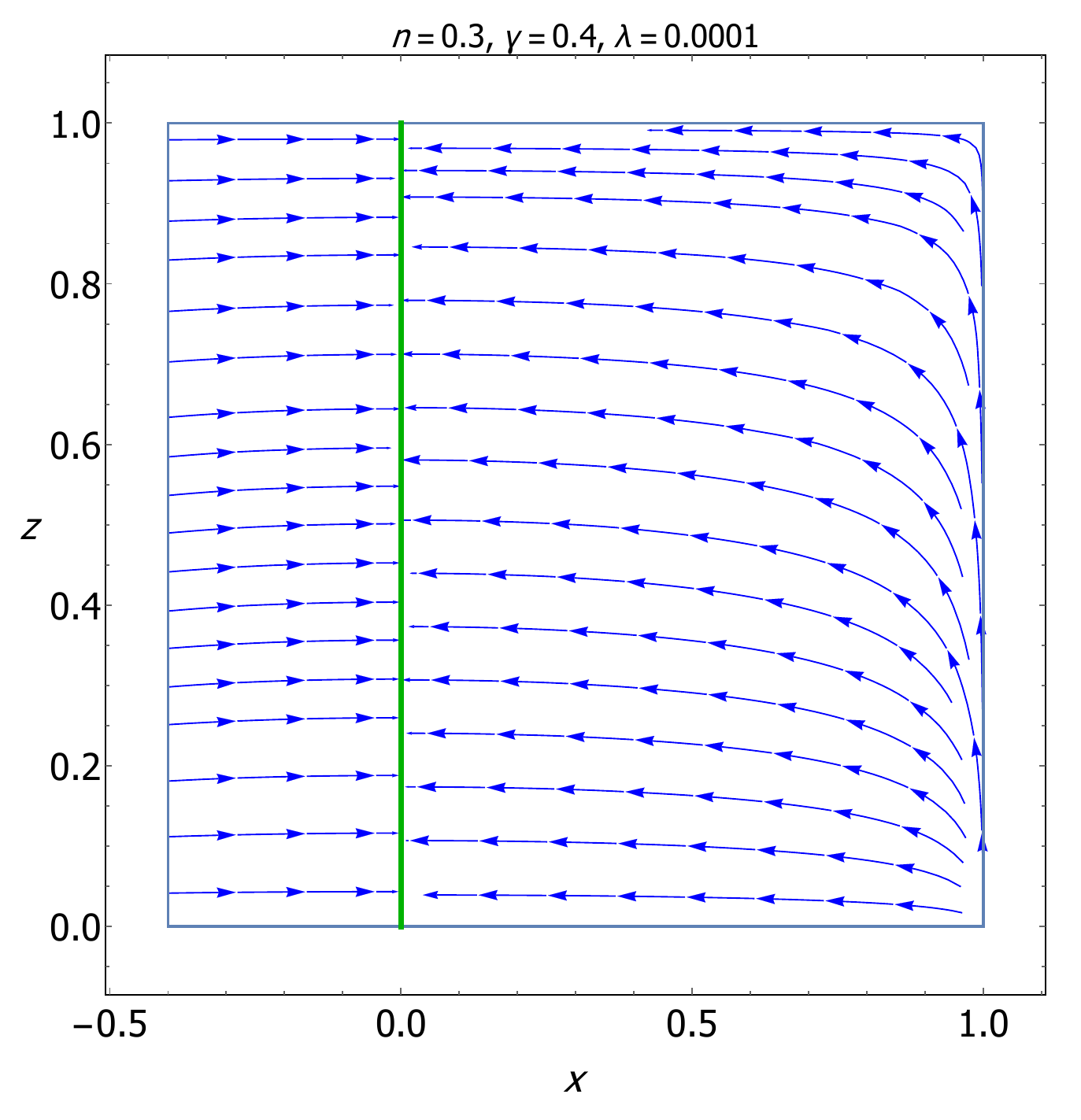}\label{P9_xz}}
	\qquad
	\subfigure[]{%
		\includegraphics[width=5.4cm,height=5.4cm]{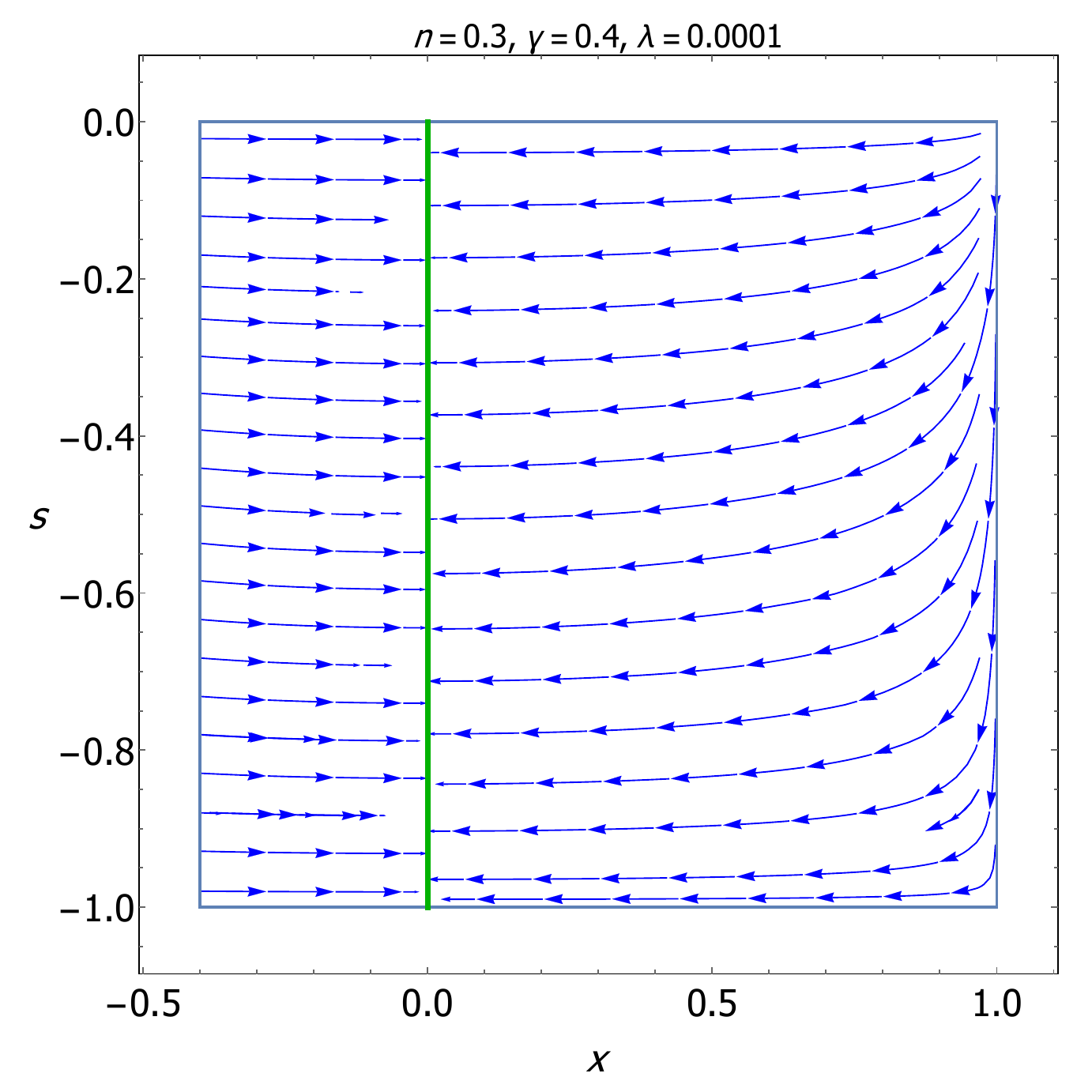}\label{P9_xs}}
		\qquad
	\subfigure[]{%
		\includegraphics[width=5.4cm,height=5.4cm]{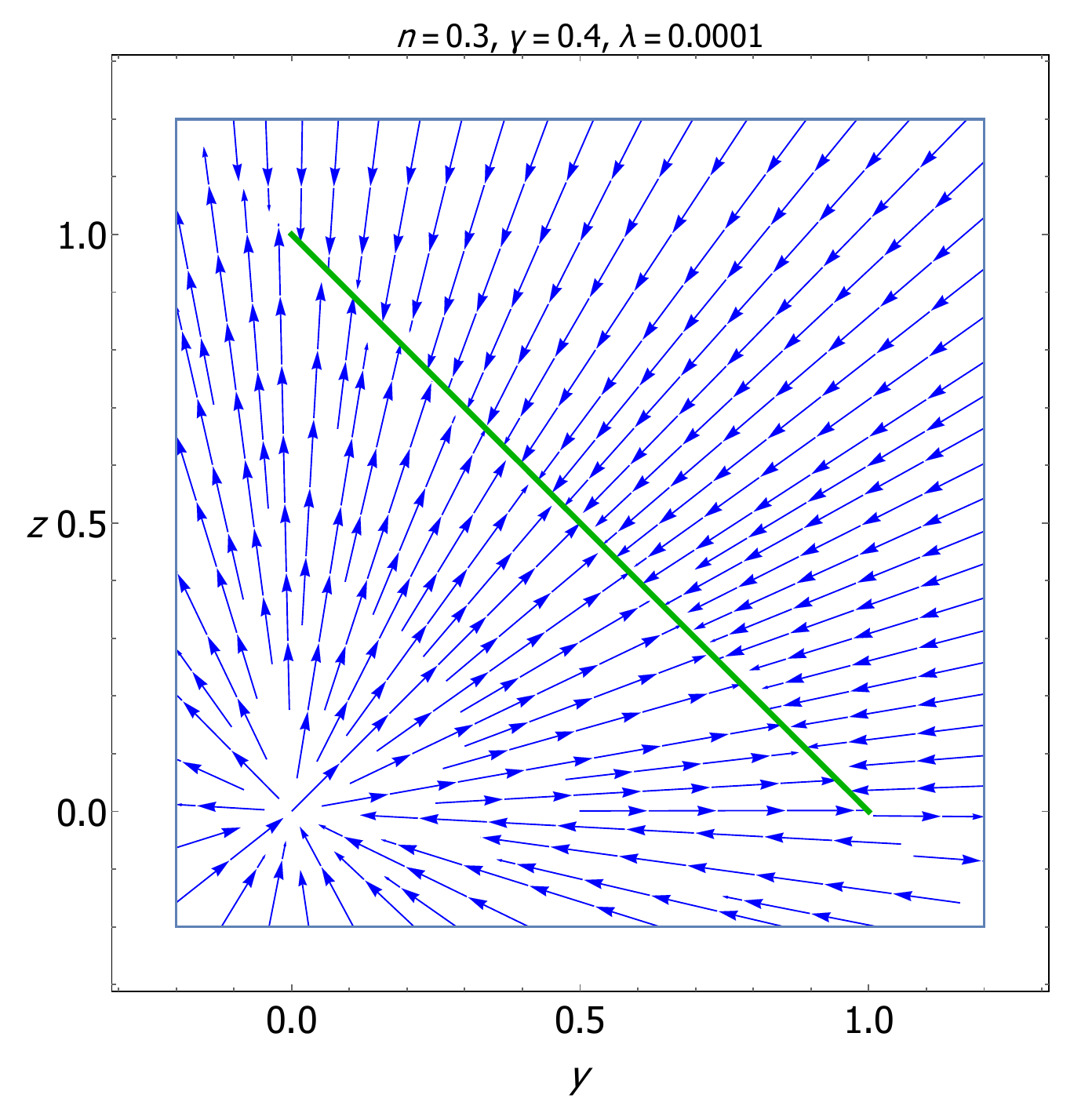}\label{P9_yz}}
	\qquad
	\subfigure[]{%
		\includegraphics[width=5.4cm,height=5.4cm]{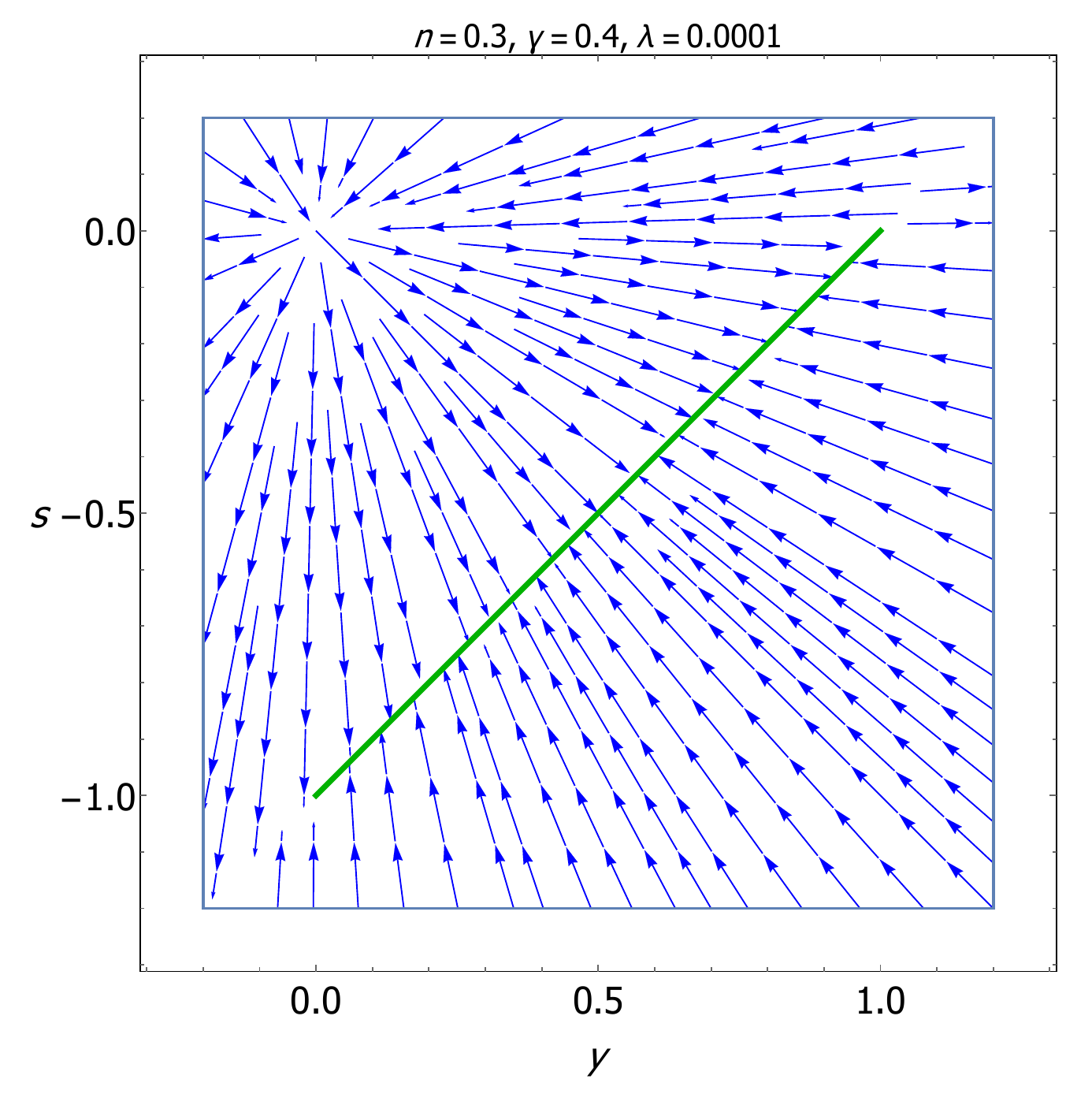}\label{P9_ys}}
		\qquad
	\subfigure[]{%
		\includegraphics[width=5.4cm,height=5.4cm]{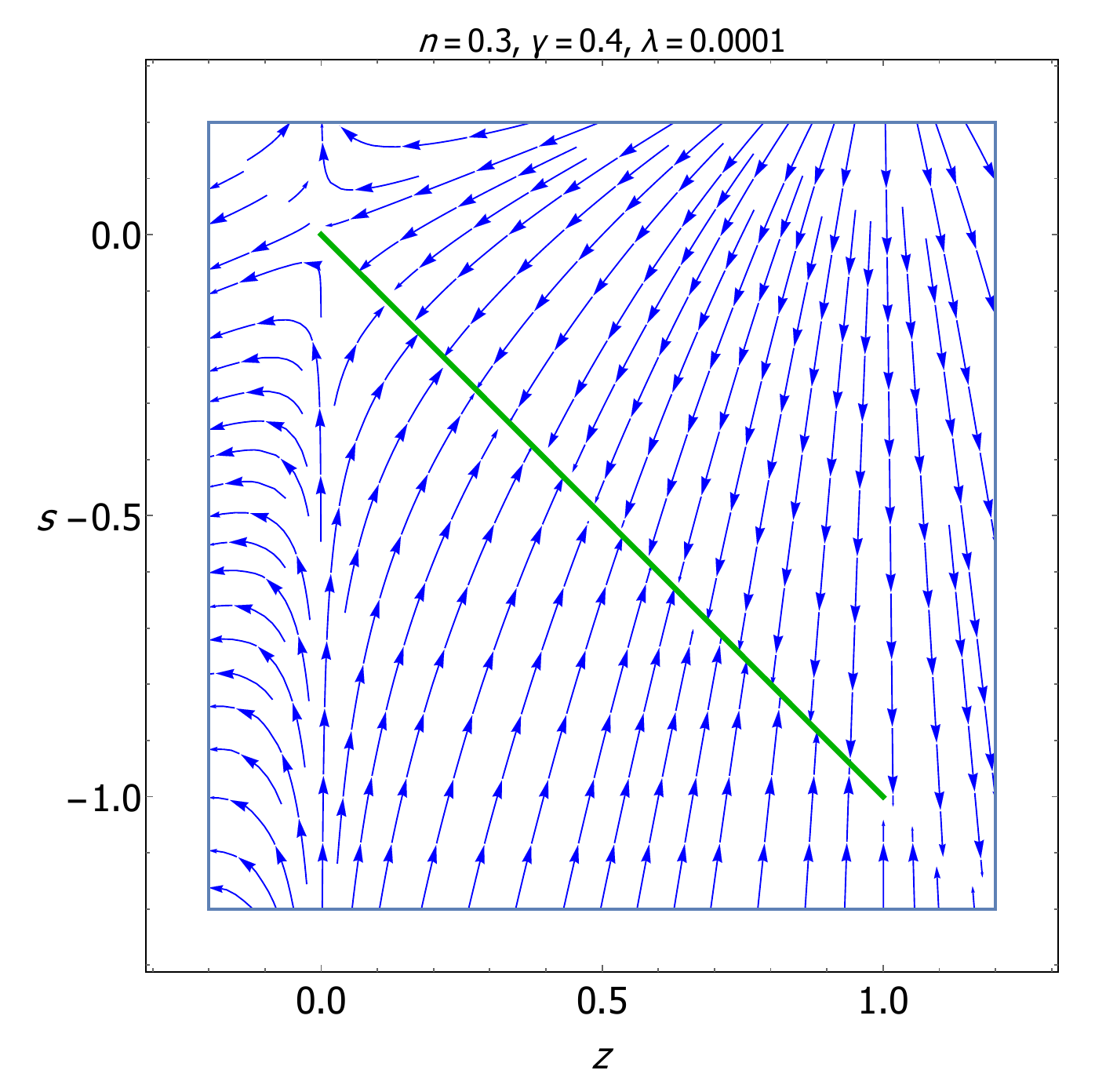}\label{P9_zs}}
	\caption{The figure shows the phase space projection of the autonomous system (\ref{autonomous_system1}) for the parameters values $n=0.3$, $\gamma=0.4$ and $\lambda=0.0001$ on the planes: x-y, x-z, x-s, y-z, y-s and z-s in panel \ref{P9_xy}, \ref{P9_xz}, \ref{P9_xs}, \ref{P9_yz}, \ref{P9_ys} and \ref{P9_zs} respectively. In all the panels the set of critical points $P_9$ is shown in the green colored line which is attractor in the phase plane.}
	\label{P9_stream}
\end{figure}


\item The sets of critical points $P_{10\pm}$ correspond to decelerated Chaplygin gas dominated solutions in the phase space. They are  non-hyperbolic sets having exactly one vanishing eigenvalue, called normally hyperbolic sets of critical points. Since all of the remaining eigenvalues are real valued and positive for $P_{10+}$ it is always repeller. But for the set of critical points $P_{10-}$ some eigen values are positive and some are negative. So, it is a saddle set. However, the set $P_{10-}$ is physically relevant solution in describing the dust dominated (here chaplygin fluid behaves as dust $\omega_{CG}=0$) decelerated phase ($q=\frac{1}{2}$). Being a saddle like solution, it shows the intermediate phase of the universe. On the other hand, the set $P_{10+}$ shows the dust dominated decelerating past attractor. For both the case the chaplygin gas behaves as dust ($\omega_{CG}=0$) which implies that $\rho_{CG}=\left( \frac{\beta}{\gamma} \right)^{\frac{1}{n+1}}$. For a specific case, $z_c =1$, the Hubble parameter will be of the form: $H=\frac{1}{\sqrt{3}} \left( \frac{\beta}{\gamma} \right)^{\frac{1}{2(n+1)}}$. 

\item The set of critical points $P_{11}$ corresponds to a decelerated Chaplygin gas dominated solution. It is  non-hyperbolic in nature having exactly one vanishing eigenvalue. So, it is also a normally hyperbolic set of critical points. The set behaves as  a repeller for $\left( -1\leq x_{c}<-\frac{1}{\sqrt{3}}~\mbox{or}~ \frac{1}{\sqrt{3}}<x_{c}\leq 1\right) $ and saddle otherwise. However, the set will be physically interested if it a saddle solution. Then it represents the dust dominated decelerated intermediate phase of the universe ($\omega_{eff}=0,~q=\frac{1}{2}$) where the chaplygin gas behaves as dust ($\omega_{CG}=0$). It is to be noted that $\omega_{CG}=0$ implies that $\rho_{CG}=\left( \frac{\beta}{\gamma} \right)^{\frac{1}{n+1}}$. Since for the existence criterion, i.e., $z_c =1$, the Hubble parameter will be of the form: $H=\frac{1}{\sqrt{3}} \left( \frac{\beta}{\gamma} \right)^{\frac{1}{2(n+1)}}$.

\end{itemize}


\subsubsection{Classical stability of the model}
\begin{figure}
	\centering
	\subfigure[]{%
		\includegraphics[width=7.9cm,height=5.5cm]{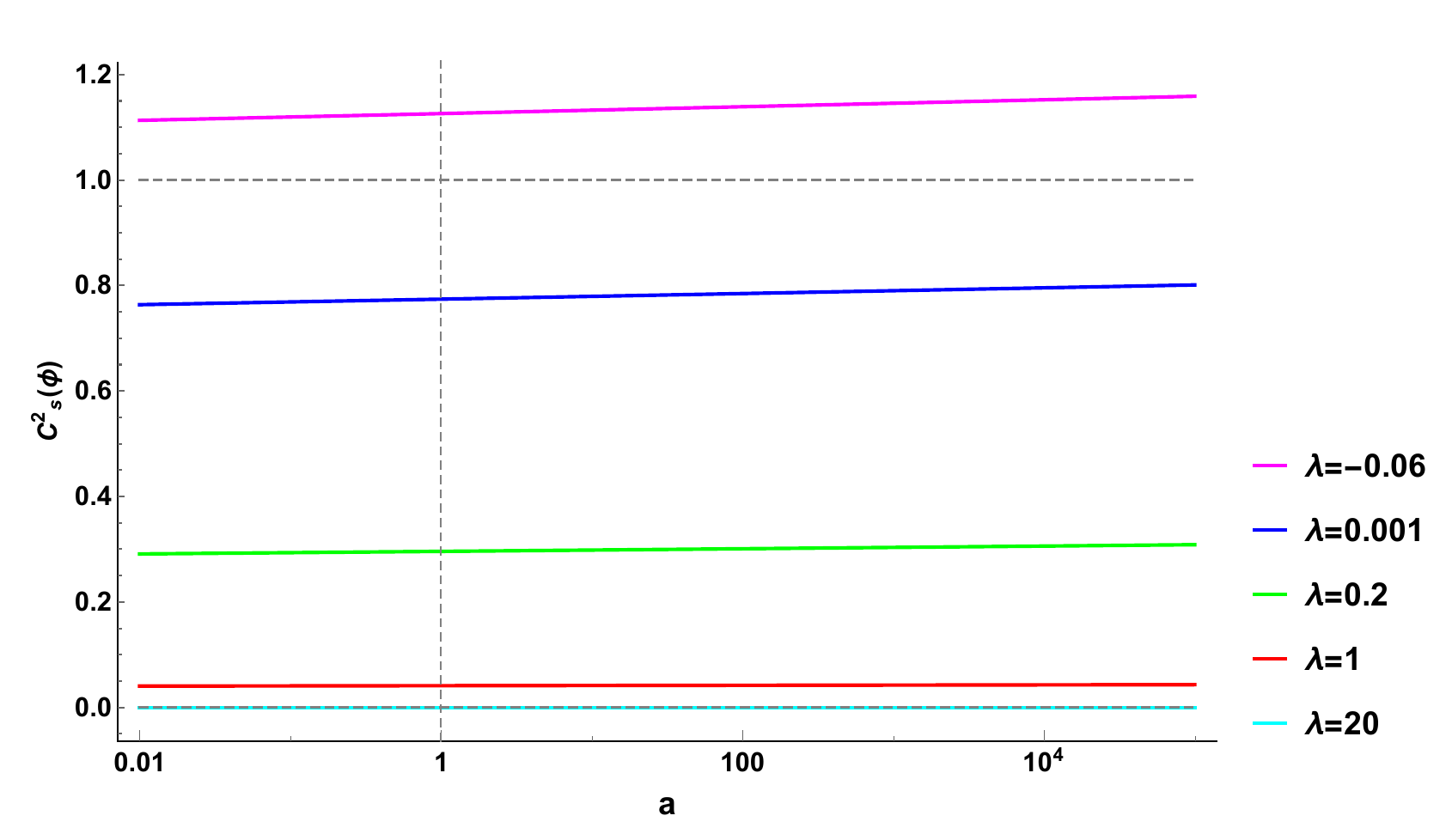}\label{SS_T}}
	\qquad
	\subfigure[]{%
		\includegraphics[width=7.9cm,height=5.5cm]{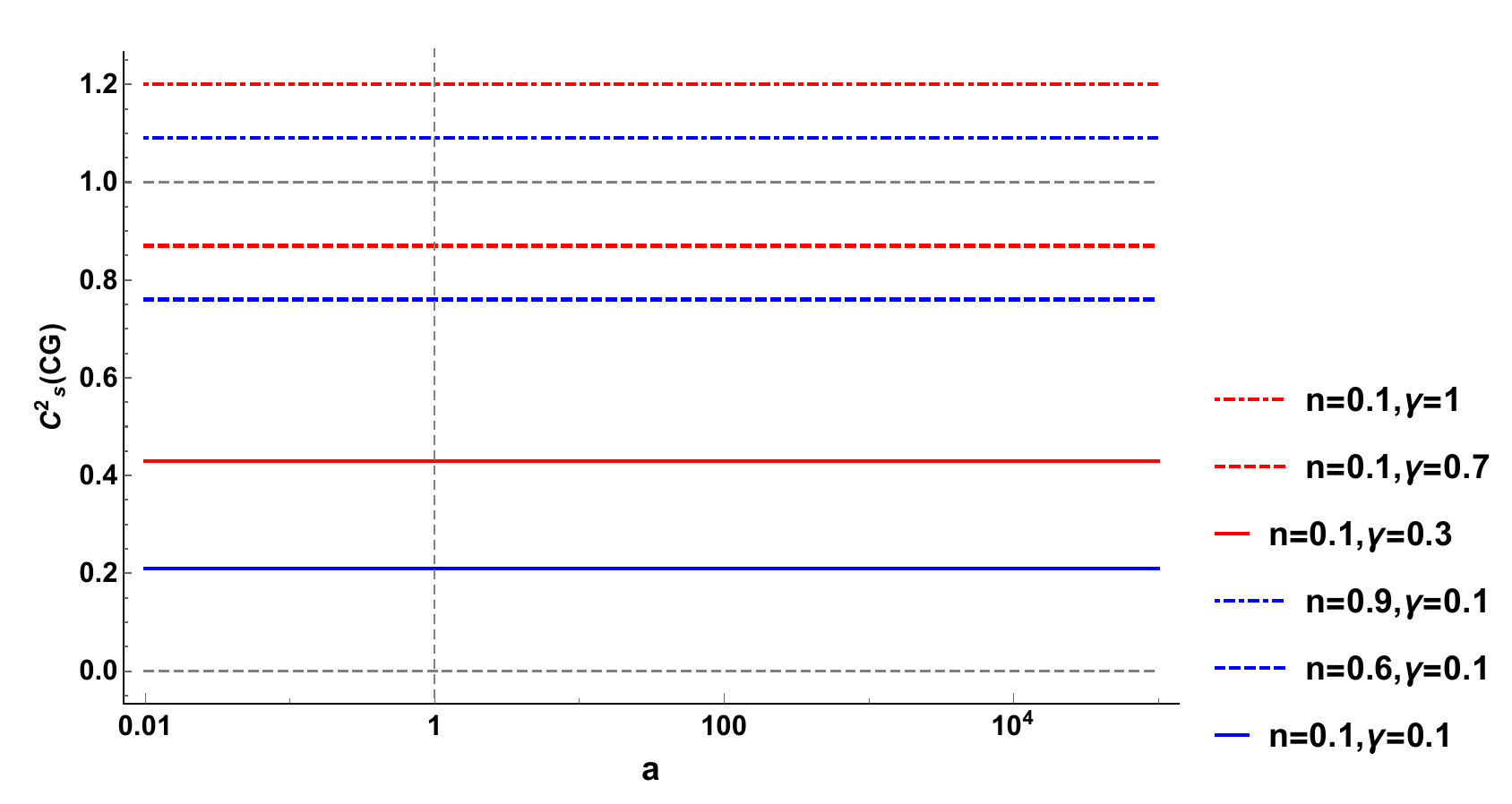}\label{SS_CG}}
	
	\caption{The figures show graphs of sound speed of tachyonic field for different choices of model parameter $\lambda$ in panel \ref{SS_T} and  for different choices of model parameters $n,\gamma$ of MCG in panel \ref{SS_CG}.}
	\label{SS}
\end{figure}

We shall now determine the stability of the models of tachyon fluid and the modified chaplygin fluid by finding the value of squared sound speed $(c_{s}^2)$. In order to obtain the stability of the model one must have the squared sound speed is greater than or equal to zero i.e., $c_{s}^2 \geq 0$, otherwise, the model is unstable. The expressions for tachyon fluid sound speed and for MCG sound speed can be evaluated in the following:
	
	\begin{equation}
		c_{s}^2(\phi)=\frac{\delta p_{\phi}}{\delta \rho_{\phi}}=(1-x^2)\left( 3+\frac{\lambda \sqrt{3y}}{x}\right) ,
	\end{equation}
	\begin{equation}
		c_{s}^2(CG)=\frac{\delta p_{CG}}{\delta \rho_{CG}}=\gamma (1+n)-\frac{n s}{z}
	\end{equation}
Therefore, from the above, one can have the stability conditions for two fluids (tachyon and MCG) as:
$(1-x^2)\left( 3+\frac{\lambda \sqrt{3y}}{x}\right)\geq 0$ and $\gamma (1+n)-\frac{n s}{z}\geq 0$. So, $c_{s}^2(\phi)$ depends on free parameters ($\lambda$) associated to tachyon fluid only. On the other hand, the value of $c_{s}^2(CG)$ will depend only on parameters $\gamma, ~ n$ associated to MCG fluid. The evolution of the squared sound speed for the models are plotted in fig. (\ref{SS}). The evolution of MCG sound speed is plotted for different MCG parameters $\gamma,~n$ in sub-fig. \ref{SS_CG} for different $\gamma$ and $n$. And it is observed that the fluid model is always stable. On the other hand, the evolution of sound speed for tachyon fluid is plotted in sub-fig. \ref{SS_T} for different model parameter $\lambda$. It is observed that the fluid model is also stable. That means they are stable independently. Therefore, considering the two fluids (tachyon and MCG) as dark components can be survived at late-times together unlike the case arrived in Ref. \cite{Noorbakhsh 2013}.


\begin{figure}
	\centering
	\subfigure[]{%
		\includegraphics[width=7.9cm,height=5.5cm]{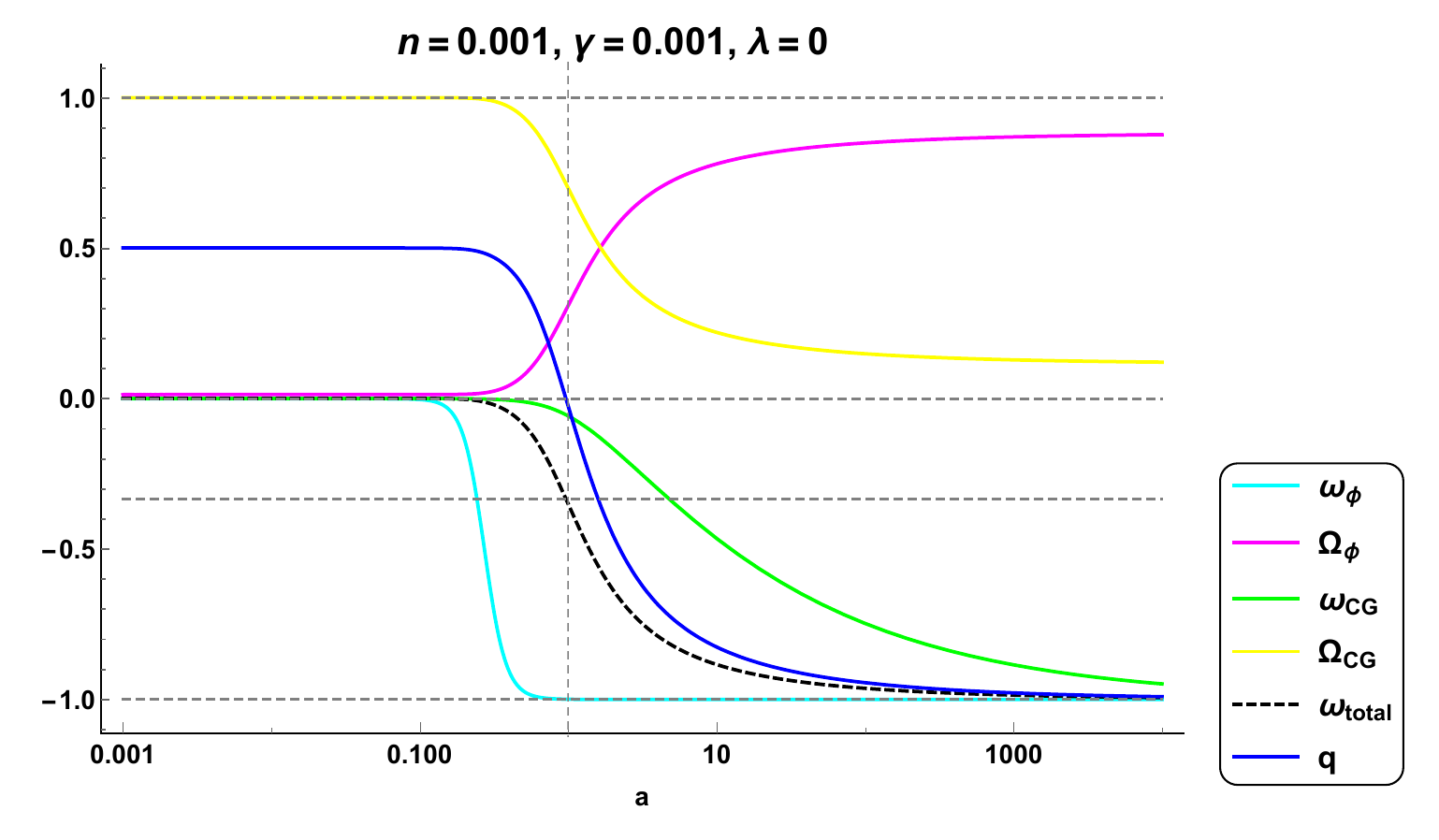}\label{fig:Evolution 1}}
	\qquad
	\subfigure[]{%
		\includegraphics[width=7.9cm,height=5.5cm]{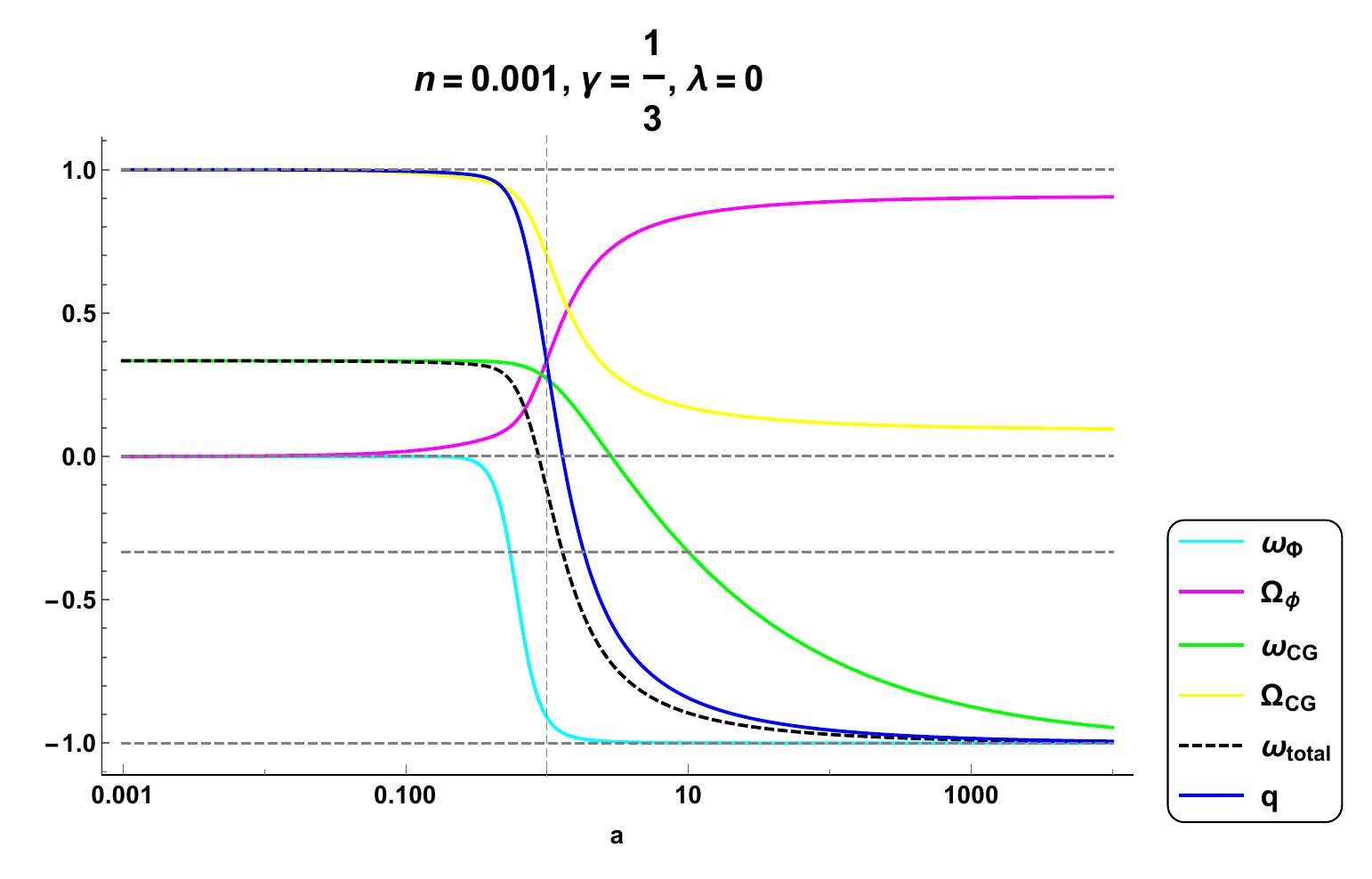}\label{fig:Evolution 2}} \\
	\qquad
	\subfigure[]{%
		\includegraphics[width=7.9cm,height=5.5cm]{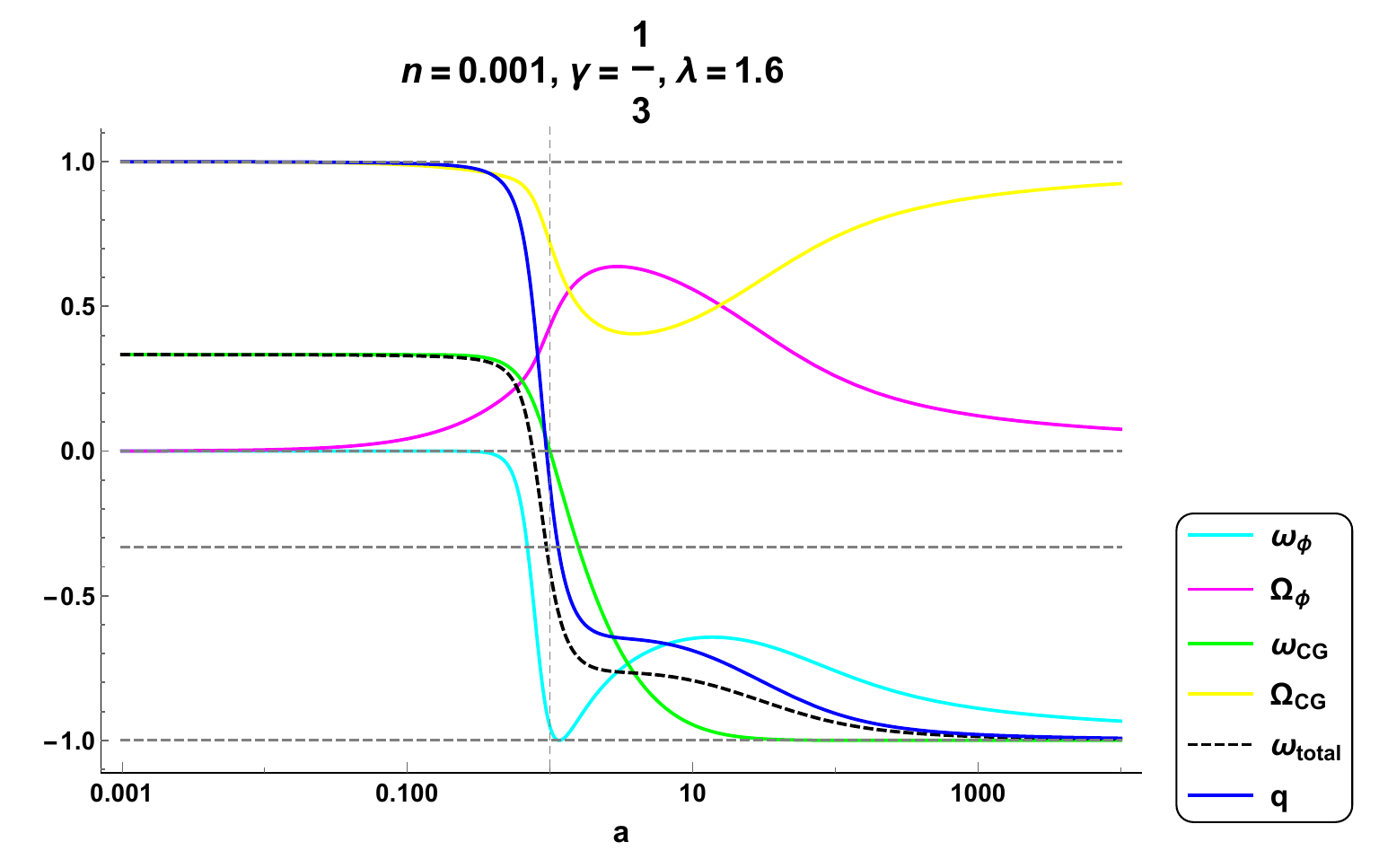}\label{fig:Evolution 3}}
	
	\caption{The figure shows the evolution of cosmological parameters for the particular values of the model parameters n, $\gamma$, $\lambda$. This shows the evolution of cosmological parameters
		where the late time evolution of the universe attracted by cosmological constant.}
	\label{Evolution}
\end{figure}

\section{Cosmological Implications of the critical points}\label{cosmological implications}
In this section, we shall discuss the cosmology of the critical points of the autonomous system (\ref{autonomous_system1}). We obtained some hyperbolic critical points and some are non-hyperbolic in nature. The linear stability theory is employed to study the stability of hyperbolic type points and numerical methods are applied to find the stability for the non-hyperbolic points. Main result are as follows:

We classify the points according to energy density contribution in the evolutionary dynamics of the universe.
The universe is completely dominated by the modified chaplygin fluid ($\Omega_{CG}=1$) represented by the critical points:  $P_1$, $P_2$, $P_{3\pm}$  $P_{4\pm}$, $P_{5\pm}$ and the sets of critical points $P_{8\pm}$, $P_{10\pm}$ and $P_{11}$. The point $P_1$ represents the accelerated evolution of the universe completely dominated by modified chaplygin gas which mimics as cosmological constant. However, it cannot describe the late-time evolution. The point may describe accelerated de Sitter solution with transient nature. The point $P_2$ is physically relevant solution only in early phase. It exhibits the early radiation dominated decelerated phase for $\gamma=\frac{1}{3}$ or dust dominated decelerated phase for $\gamma=0$ where the chaplygin gas behaves as radiation and dust respectively. The points $P_{3\pm}$ have similar behaviour to the point $P_1$. The points $P_{4\pm}$ and $P_{5\pm}$ have the similar evolutionary scheme to that of the point $P_2$. Although the accelerated evolution of the universe is achieved by the set of points $P_{8\pm}$, these do not  describe the late-time evolution. In fact, the points can describe the transient de Sitter phase of the universe. The intermediate phase of the universe is described by the sets $P_{10\pm}$. Here, the universe evolves in dust era where the chaplygin gas mimics as dust. The set $P_{11}$ has the similar behaviour of evolution with that of the sets $P_{10\pm}$.\\

Next, the tachyon fluid dominated solutions are achieved by the sets of points $P_6$ and $P_7$. From numerical investigation, we observe that the set $P_6$ has not capable of providing late-time evolution due to its unstable in phase space. On the other hand, the set of points $P_7$ is physically interested set in late-times see in fig. (\ref{P7-perturbation}). Late-time accelerated evolution of the universe is observed by the set for  $y_{c} \sqrt{1-x_{c}^2}> s_{c}+\frac{1}{3}$, $s_{c}>0$. And depending upon some parameters the set can depict the late-time accelerated universe evolving in quintessence era only where the tachyon field behaves as quintessence fluid.\\

Finally, a tachyon-Chaplygin fluid scaling solution is achieved by the set of critical points $P_9$. Interestingly, both the fluids behave as cosmological constant here ($\omega_{\phi}=-1,~\omega_{CG}=-1$). The ratio of the energy densities is $\frac{\Omega_{CG}}{\Omega_{\phi}}=\frac{z_c}{1-z_c}$which provides the possible mechanism for alleviating the coincidence problem. This is a normally hyperbolic set and stability of this set can be found by evaluating the signature of non-vanishing eigenvalues. The fig.(\ref{P9-perturbation}) shows its stability in phase space. For $z_{c}\longrightarrow 1$, the accelerated universe is dominated by chaplygin gas and attracted in cosmological constant era ($\Omega_{CG}\longrightarrow 1, ~\Omega_{\phi}\longrightarrow 0, ~\omega_{eff}=-1,~ q=-1$). On the other hand, for $z_{c}\longrightarrow 0$, the universe will evolve in tachyonic fluid dominated era ($\Omega_{\phi}\longrightarrow1,~\Omega_{CG}\longrightarrow 0 $). \\

In conclusion, our study reveals that the solutions dominated by the modified chaplygin gas cannot provide the late-time acceleration of the universe. These solutions are relevant only in early evolution of the universe. The fig. \ref{fig:Evolution 1} for parameter values ($n,~\gamma,~\lambda$)=($0.001,~0.001,~0$) exhibits that the universe evolves first in MCG dominated decelerated era then it enters into the tachyon fluid dominated phase attracted in $\Lambda$CDM model ($\omega_{eff}=-1$) where the tachyon fluid EoS evolves as cosmological constant. Here, the equation of state parameter $\omega_{CG}$ evolves from dust like fluid in early times and in future it evolves like quintessence. The fig. \ref{fig:Evolution 2} for model parameters ($n,~\gamma,~\lambda$)=($0.001,~\frac{1}{3},~0$) It is evident that the early evolution of the universe is achieved in radiation era ($\omega_{eff}=0.33$) which is completely dominated by the MCG fluid where it behaves as radiation ($\omega_{CG}$=0.33) and the epoch is decelerated era. In future, the universe enters into the tachyon fluid domination era where acceleration is driven by tachyon mimicking as cosmological constant.
Finally the fig. \ref{fig:Evolution 3} describes the late-time evolution of the universe dominated by the MCG fluid. This is possible only due to the existence of the scaling solution $P_9$. In this figure for parameter values ($n,~\gamma,~\lambda$)=($0.001,~\frac{1}{3},~1.6$) and with initial conditions: $x(0)=0.22,~y(0)=0.42,~z(0)=0.72,~s(0)=0.001$ it can be translated that late-time accelerated evolution of universe is dominated by the MCG fluid with cosmological parameters for this scenario are: $\Omega_{\phi}=0.28,~\Omega_{CG}=0.72,~\omega_{\phi}=-1,~\omega_{CG}=-1,~\omega_{eff}=-1,~q=-1$. Here, the MCG and the tachyon behave as cosmological constant like fluid. Interestingly, from the all subfigures of fig. (\ref{Evolution}) it is evident the late-time accelerated evolution of the universe is attracted in $\Lambda$CDM era.


\section{Cosmography of the model}\label{cosmography}
In this section, we shall investigate the cosmographic analysis of our model. Sahni et al. in Ref. \cite{Sahni 2003} and Alam et al. in Ref. \cite{Alam 2003} first proposed two geometrical diagnostic parameters $r$ and $s$, called state finder parameters which are defined as follows:
\begin{equation}\label{r state}
	r\equiv \frac{1}{a H^{3}} \frac{d^3a}{dt^3},
\end{equation}

\begin{equation}\label{s state}
	s\equiv\frac{r-1}{3(q-\frac{1}{2})}
\end{equation}
where $a$ is the scale factor and $q=-\frac{a }{a^{2}}\frac{d^2a}{dt^2}$ is the deceleration parameter. These two geometrical (since they depend only on scale factor) dimensionless parameters are used to diagnose the properties of various dark energy models and to compare / discriminate them with the $\Lambda$CDM in a model independent manner. Note that the values of two parameters ($r,~s$)=($1,~0$) corresponds to the $\Lambda$CDM model while the values ($r,~s$)=($1,~1$) describes the standard cold dark matter model. This geometric diagnosis on the dark energy models has been further extended by evaluating the Taylor series expansion in respect of scale factor ($a$) about the present time and thus, one can have some more general geometrical model independent parameters denoted by $j$ (jerk), $s$ (snap), and $l$ (lerk) which are called the Cosmographic Parameters (CPs) \cite{Visser 2004,Visser 2005,Pan 2014,Singh 2020} and they are defined as follows:
\begin{equation}\label{CPs}
	j\equiv \frac{1}{a H^{3}}\frac{d^3a}{dt^3},~s\equiv \frac{1}{a H^{4}}\frac{d^4a}{dt^4},~l\equiv \frac{1}{a H^{5}}\frac{d^5a}{dt^5},
\end{equation}
Note that this snap $s$ is not same with the parameter $s$ defined in Eqn (\ref{s state}). But the jerk $j$ is same as $r$ defined in Eqn (\ref{r state}). Now, for our model, we compute the statefinder parameters in the following.

First, the energy conservation equation in (\ref{continuity}) gives the energy density for tachyonic field:
\begin{equation}\label{energy_density_T}
	\rho_\phi=\rho_{\phi0}~ a^{-3(1+\omega_{\phi})}
\end{equation}
and the energy conservation equation (\ref{continuity_CG}) for MCG provides the energy density for MCG as
\begin{equation}\label{energy_density_MCG}
	\rho_{CG}=\left\lbrace \left(\frac{\beta}{\gamma+1} \right) \left( 1-\frac{1}{a^{3(\gamma+1)(n+1)}} \right)+\left(\frac{\rho_{CG0}}{a^{3(\gamma+1)}} \right)^{n+1}    \right\rbrace^{\frac{1}{n+1}} 
\end{equation}
where the scale factor has been normalized to the present value $a(t_0)=1$, and $\rho_{\phi0}$ and $\rho_{CG0}$ are present-time energy densities
for tachyonic field and MCG respectively. It should be mentioned that the equation of state of LIVE, corresponding to a cosmological constant, is
$p_{CG}=-\rho_{CG}$. Inserting this in Eqn (\ref{MCG EoS 1}) gives $\rho_{CG}=\left(\frac{\beta}{1+\gamma} \right)^{\frac{1}{n+1}}$. This appears as the final density for large values of the scale factor (as given in Eqn.(\ref{continuity_CG})). Thus, the Eqn. (\ref{energy_density_MCG}) shows that the behaviour of the chaplygin gas approaches that of LIVE with a constant
density as given in P6L51-P10L50, which may be represented by a cosmological constant.\\

Using equations (\ref{energy_density_T}), (\ref{energy_density_MCG}) the Friedmann equation (\ref{Friedmann}) gives the expression for Hubble parameter in terms of redshift(z) as:
\begin{equation}\label{E}
	E(z)=\frac{H(z)}{H_{0}}=\left[ \Omega_{\phi0}~ (1+z)^{3(1+\omega_{\phi})}+ \left\lbrace \left( \frac{1}{3 H_{0}^{2}}\right)^{n+1}  \left(\frac{\beta}{\gamma+1} \right) \left( 1-(1+z)^{3(\gamma+1)(n+1)} \right)+\left( \Omega_{CG0} (1+z)^{3(\gamma+1)} \right)^{n+1}    \right\rbrace^{\frac{1}{n+1}}  \right]^{\frac{1}{2}} 
\end{equation}
where $H_{0}$ is the present value of the Hubble parameter and scale factor $a=\frac{1}{1+z}$.

\begin{figure}
	\centering
	\subfigure[]{%
		\includegraphics[width=8.4cm,height=8.4cm]{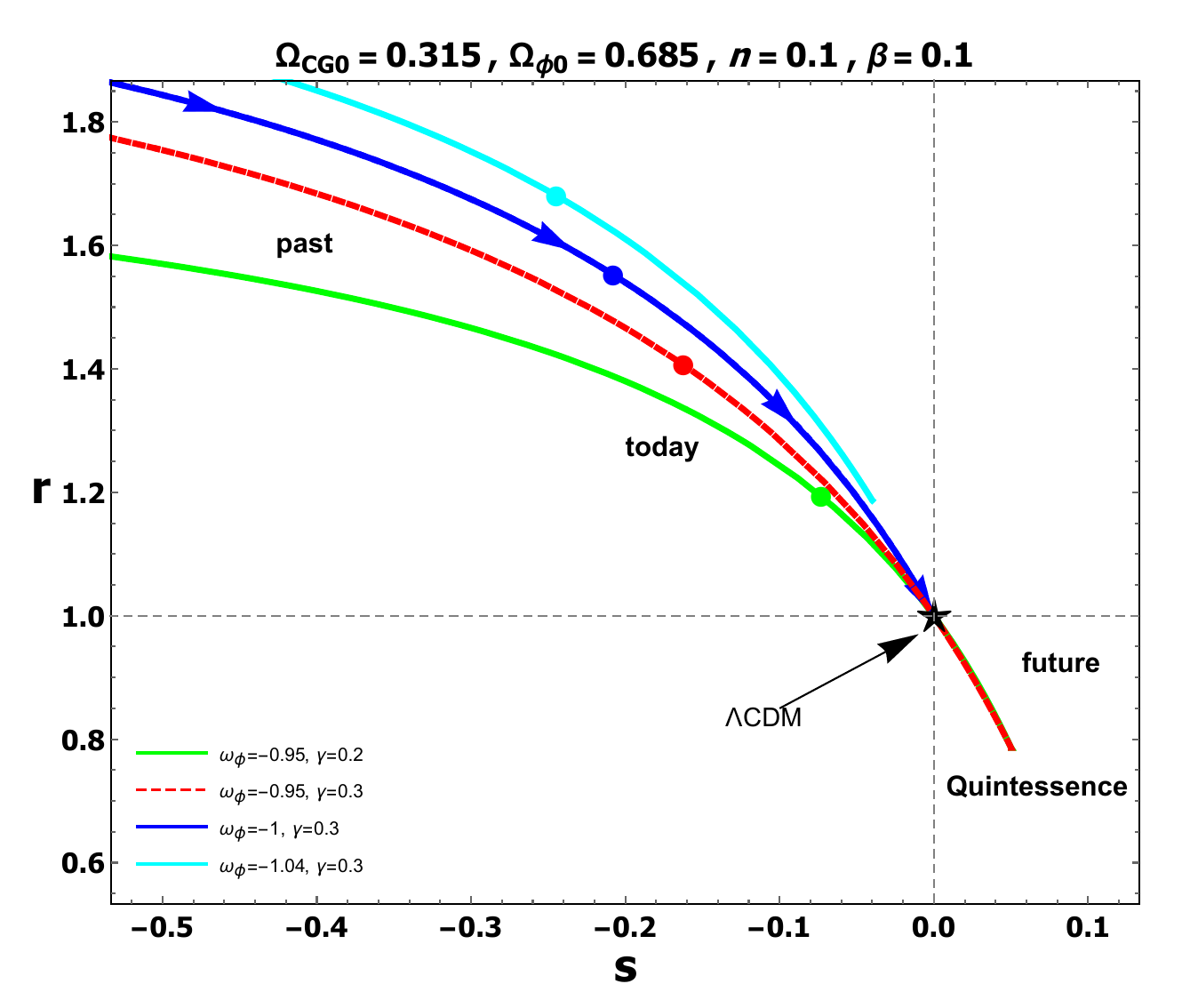}\label{sr}}
	\qquad
	\subfigure[]{%
		\includegraphics[width=8.4cm,height=8.4cm]{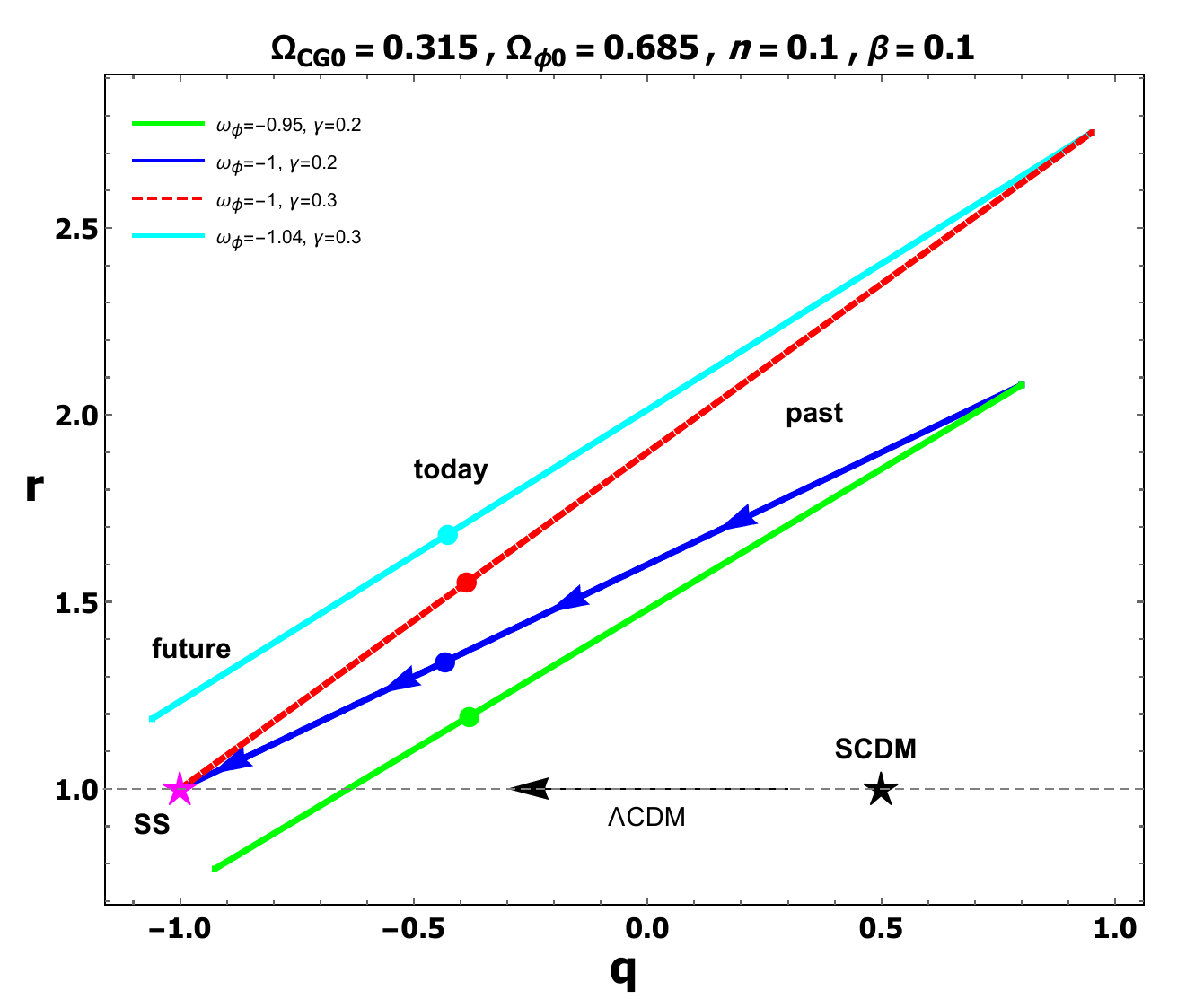}\label{qr}}
	
	\caption{ The figures show different time evolution trajectories of the state finder pair $(s ,r)$ in panel \ref{sr} and $(q,r)$ in panel \ref{qr} for different values of parameters as indicated in each panel. The coloured circles denote the present value of the state finder parameter $(s_{0},r_{0})$ in panel \ref{sr} and $(q_{0},r_{0})$ in panel \ref{qr}. In this figure the present value of the Hubble parameter $H_{0}=69~kms^{-1}Mpc^{-1}$. }
	\label{State finder parameters}
\end{figure}

For convenience, we introduce the dimensionless Hubble rate $E(z)\equiv H(z)/H_{0}$, where $H_{0}$ is the Hubble constant. Now, the parameters $\left\lbrace q,r,s\right\rbrace $ in Eqns (\ref{r state}) and (\ref{s state}) can be computed in terms of $E(z)$ as follows:
\begin{equation}
	q(z)=-1+(1+z) \frac{E_{z}(z)}{E(z)},
\end{equation}
\begin{equation}\label{q-transit}
q(z)=\frac{(z+1) \left(\frac{3^{-n} (z+1)^{3 \gamma +3 (\gamma +1) n+2} \left((\gamma +1) H_{0}^2 3^{n+1} \Omega_{CG0}^{n+1}-\beta  \left(\frac{1}{H_{0}^2}\right)^n\right) \left(\Delta\right)^{\frac{1}{n+1}-1}}{H_{0}^2}+3 \Omega_{\phi0} (\omega_{\phi} +1) (z+1)^{3 \omega_{\phi} +2}\right)}{2 \left(\left( \Delta \right)^{\frac{1}{n+1}}+\Omega_{\phi0} (z+1)^{3 \omega_{\phi} +3}\right)}-1,
\end{equation}
where $\Delta=\left(\Omega_{CG0} (z+1)^{3 \gamma +3}\right)^{n+1}-\frac{\beta  3^{-n-1} \left(\frac{1}{H_{0}^2}\right)^{n+1} \left((z+1)^{3 (\gamma +1) (n+1)}-1\right)}{\gamma +1}$
\begin{equation}
	r(z)=q(z)(1+2q(z))+(1+z)q_{z}(z)
\end{equation}
Furthermore $s(z)$ is given by (\ref{s state}), and suffix $z$ stands for derivative with respect to $z$. \\
The trajectories of statefinder parameters $\left\lbrace r,~s\right\rbrace $ and $\left\lbrace r,~q\right\rbrace $ are shown in the fig. (\ref{State finder parameters}) for different values of $\omega_{\phi}$ and $\gamma$ with the values of model parameters:
$n=0.1$, $\beta=0.1$ and the initial values of cosmological parameters: $(\Omega_{CG0},~\Omega_{\phi0})=(0.315,~0.685)$ where the present value of the Hubble parameter is taken as $H_0= 69$ km $s^{-1}$ Mp$c^{-1}$. The arrows indicate the direction of evolution of the trajectories. 

The time evolution trajectories of the state finder pair $(s ,r)$ are shown in panel \ref{sr} and the trajectories for the pair $(q,r)$ are shown in panel \ref{qr}. The coloured circles denote the present value of the state finder parameter $(s_{0},r_{0})$ in panel \ref{sr} and $(q_{0},r_{0})$ in panel \ref{qr}. The sub. fig. \ref{sr} shows that trajectories started from the past (in region $r>1$ and $s<-0.5$) crossing the $\Lambda$CDM at $s=0,~r=1$ then tends to quintessence region in future for $\omega_{\phi}>-1$ and the trajectory tends to $\Lambda$CDM in future for $\omega_{\phi}=-1$. Finally, for $\omega_{\phi}<-1$, the evolution will go towards the $\Lambda$CDM but not reach at $\Lambda$CDM. On the other hand, the sub fig.\ref{qr} shows the phase transition of deceleration parameter $q$ and the trajectories finally enter to the steady state (SS) region for $\omega_{\phi}=-1$ and the line $r=1$ represents the $\Lambda$CDM model.   \\

Now, the most general cosmographic parameters: jerk($j$), snap($s$), and lerk($l$) are defined in Eqn (\ref{CPs}) which are computed for our model in terms of $E(z)$ in the following:
\begin{equation}
	s(z)=-\left\lbrace r(z)(2+3q(z))+(1+z)r_{z}(z) \right\rbrace ,
\end{equation}
\begin{equation}
	l(z)=-\left\lbrace s(z)(3+4q(z))+(1+z)s_{z}(z) \right\rbrace,
\end{equation}
\begin{figure}
	\centering
		\centering
	\subfigure[]{%
		\includegraphics[width=8.4cm,height=8.4cm]{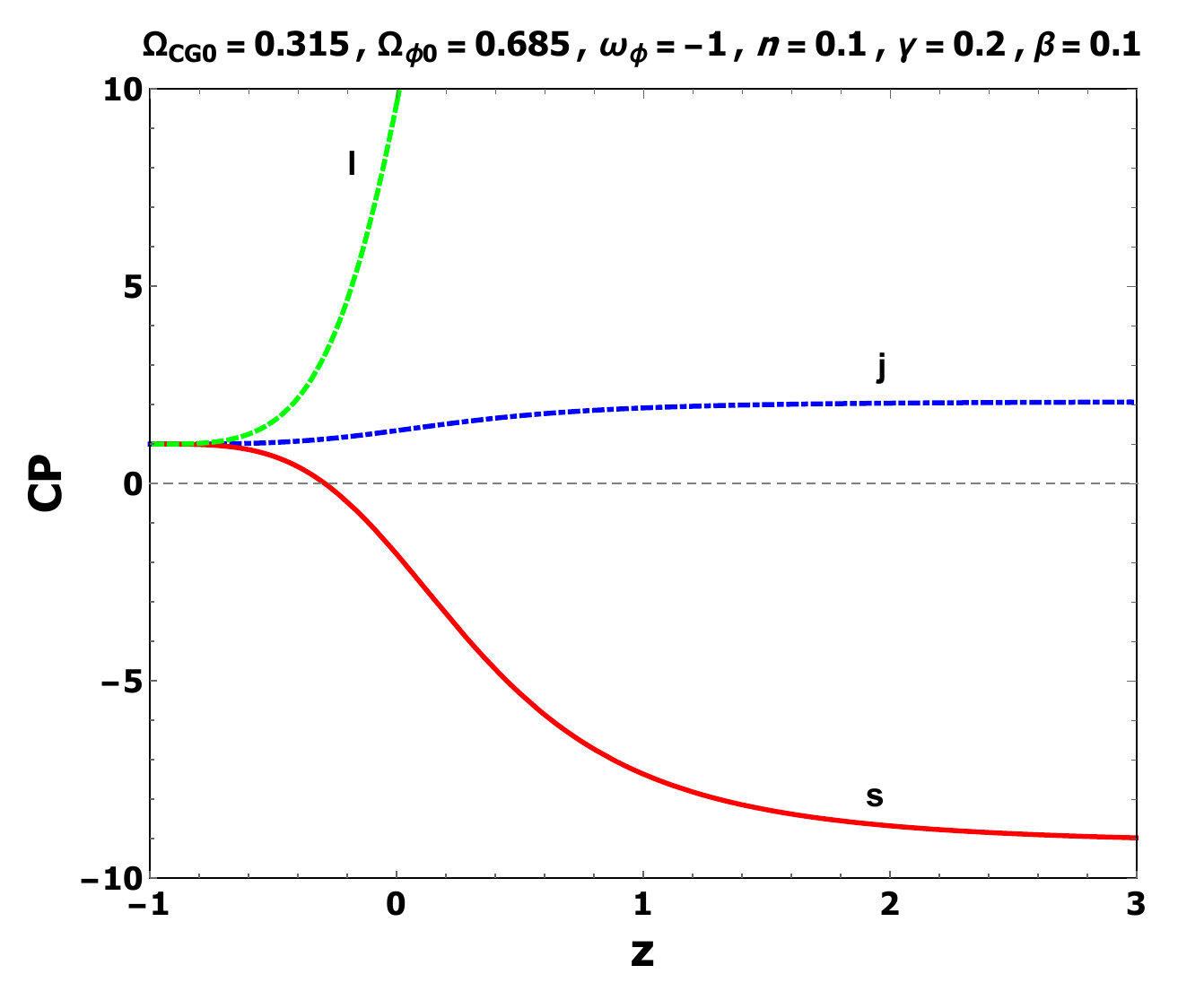}\label{CPs}}
	\qquad
	\subfigure[]{%
		\includegraphics[width=8.4cm,height=8.4cm]{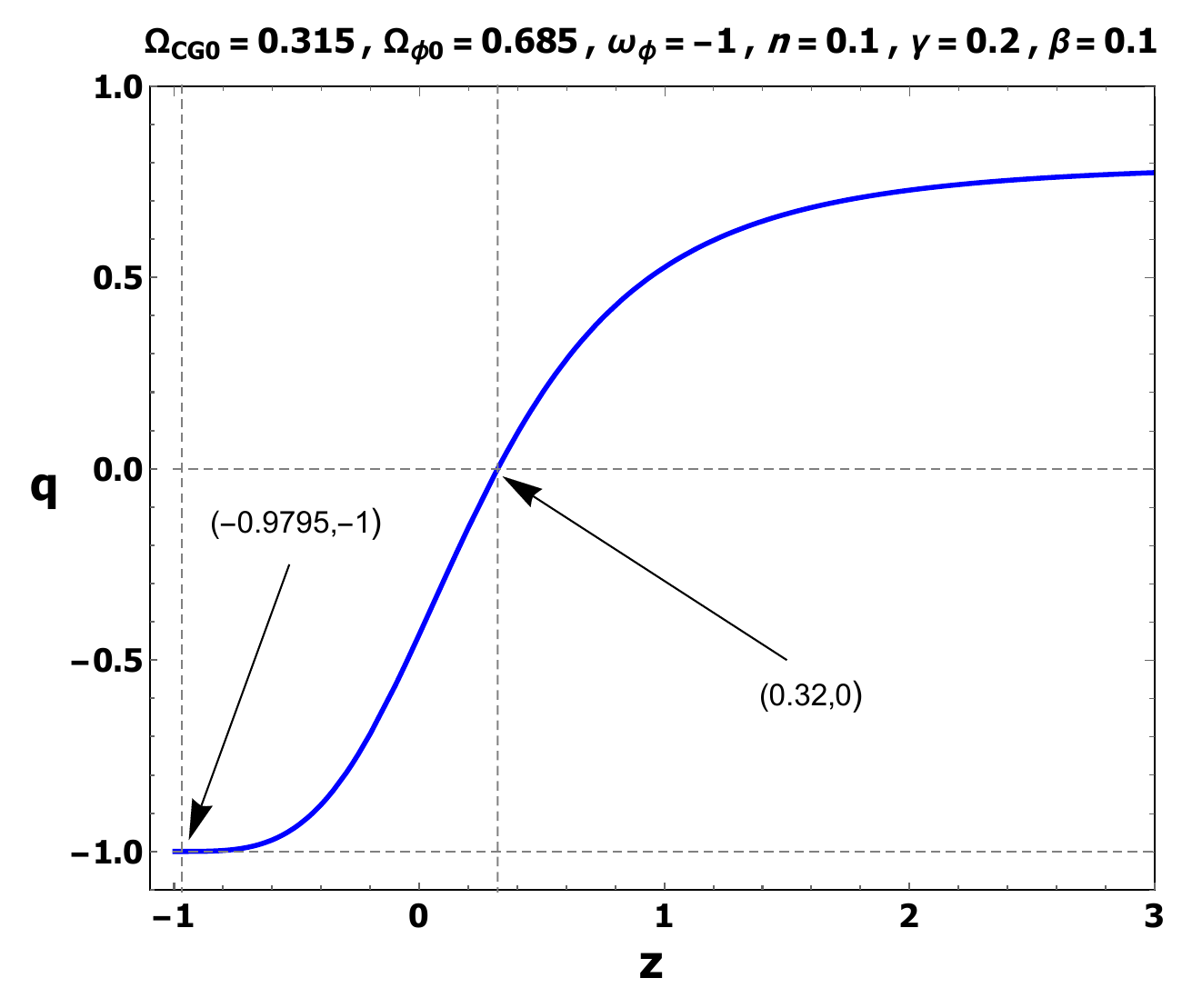}\label{zq}}
	
	\caption{The figure shows the trajectories of cosmographic parameters $j,s~\mbox{and}~l$ in panel \ref{CPs} and the deceleration parameter $q$ is shown in panel \ref{zq} as a function of redshift $z$. Panel \ref{zq} shows phase transition from decelerating to accelerating at $z=0.32$ and there will be a transition to a $\Lambda$CDM-era at $z=-0.9795$. In this figure the present value of the Hubble parameter $H_{0}=69~kms^{-1}Mpc^{-1}$. }
	\label{CP}
\end{figure}
The fig. (\ref{CP}) exhibits the trajectories of CPs: jerk, snap and lerk parameters for the values of model parameters ($n,~\gamma,~\beta$)=(0.1,~0.2,~0.1) with the initial values of the cosmological parameters: ($\Omega_{CG0},~\Omega_{\phi0},~\omega_{\phi}$)=($0.315,~0.685,~-1$) and the present value of the Hubble parameter is taken as $H_{0}=69~kms^{-1}Mpc^{-1}$.
From the figure, it is shown that the parameter $s$ changes its sign from negative to positive at low red shift i.e., when $z\longrightarrow -1$ and the parameter $l$ sharply decreases its value when the evolution goes to the low red shift region. It is worthy to note that the parameters $j$ and $l$ will remain positive for their entire evolution. In fact, all the parameters will converge to the value 1 at late-time (as $z\longrightarrow-1$).  Therefore, our model can predict the $\Lambda$CDM in its future evolution. The plot of deceleration parameter $q$ (given in Eqn.(\ref{q-transit})) against redshift $z$ in the sub-fig. \ref{zq} exhibits the phase transition of the evolution of the universe from decelerating to accelerating at $z=0.32$ and there will be an ultimate transition to a $\Lambda$CDM-era at $z=-0.9795$.\\
\subsection{$O_m$ diagnostic}
$O_m$ diagnosis \cite{Sahni 2008} is another model independent test to discriminate the dark energy models from $\Lambda$CDM. It depends only on the Hubble parameter and it is defined by \cite{Singh 2020}
\begin{equation}
	Om(z)=\frac{\frac{H^2(z)}{H_{0}^2}-1}{(1+z)^3-1}
\end{equation}
For our model, by using (\ref{E}), we can obtain $Om$ diagnostic parameter as 
\begin{equation}
	Om(z)=\frac{\left[ \left\lbrace \left(\Omega_{CG0} (z+1)^{3 (\gamma +1)}\right)^{n+1}+\frac{\beta  3^{-n-1} \left(\frac{1}{H_{0}^2}\right)^{n+1} \left(1-(z+1)^{3 (\gamma +1) (n+1)}\right)}{\gamma +1}\right\rbrace ^{\frac{1}{n+1}}+\Omega_{\phi0} (z+1)^{3 (\omega_{\phi} +1)}\right] ^2-1}{(z+1)^3-1}
\end{equation}

\begin{figure}
	\centering
	\includegraphics[width=8cm,height=8cm]{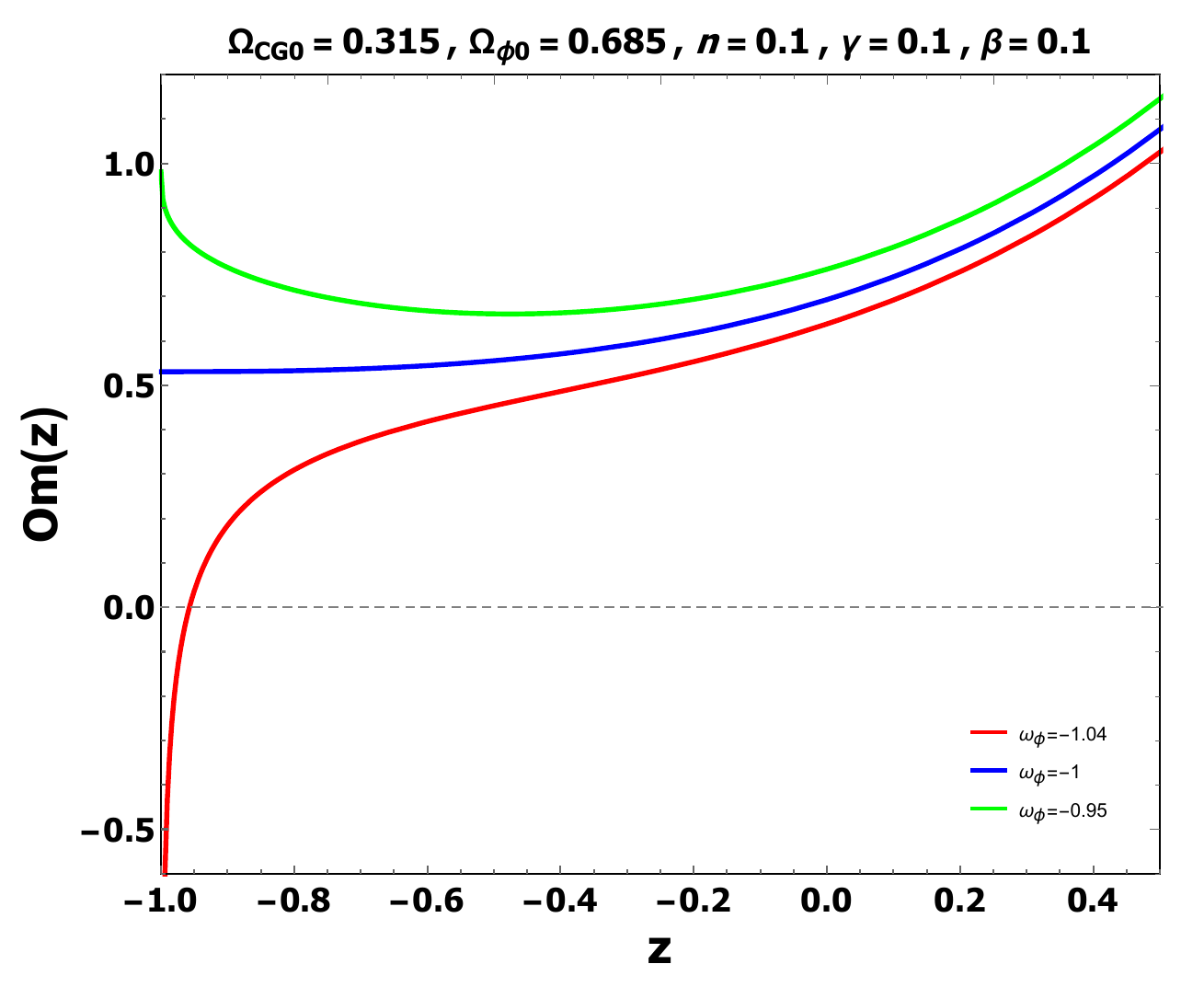}
	
	\caption{The figure shows $O_m-z$ trajectories. In this figure the present value of the Hubble parameter $H_{0}=69~kms^{-1}Mpc^{-1}$. }
	\label{Om}
\end{figure}
Note that the constant behavior of $O_m$ implies that the DE can behave as cosmological constant and positive slope of $O_m$ indicates that the DE describes the phantom like fluid. Finally, the negative slope of $O_m$ provides that the DE behaves as quintessence. We now show the trajectories of $O_m-z$ diagnostic for different values of $\omega_{\phi}$ against the redshift ($z$) in the fig. (\ref{Om}) with the model parameters: ($n,~\gamma,~\beta$)=(0.1,~0.1,~0.1) and the cosmological parameters: ($\Omega_{CG0},~\Omega_{\phi0}$)=($0.315,~0.685$)where the present value of the Hubble parameter is taken as: $H_{0}=69~kms^{-1}Mpc^{-1}$. The fig shows that the positive slope of trajectory will occur for $\omega_{\phi}<-1$ which indicates that our model will mimic as phantom in future. The trajectory for $\omega_{\phi}>-1$ shows the negative slope mimicking the quintessence model in future and the $\Lambda CDM$ model can be observed in future when $\omega_{\phi}$ takes the value $\omega_{\phi}=-1$. 

\begin{figure}
	\centering
	\includegraphics[width=8cm,height=8cm]{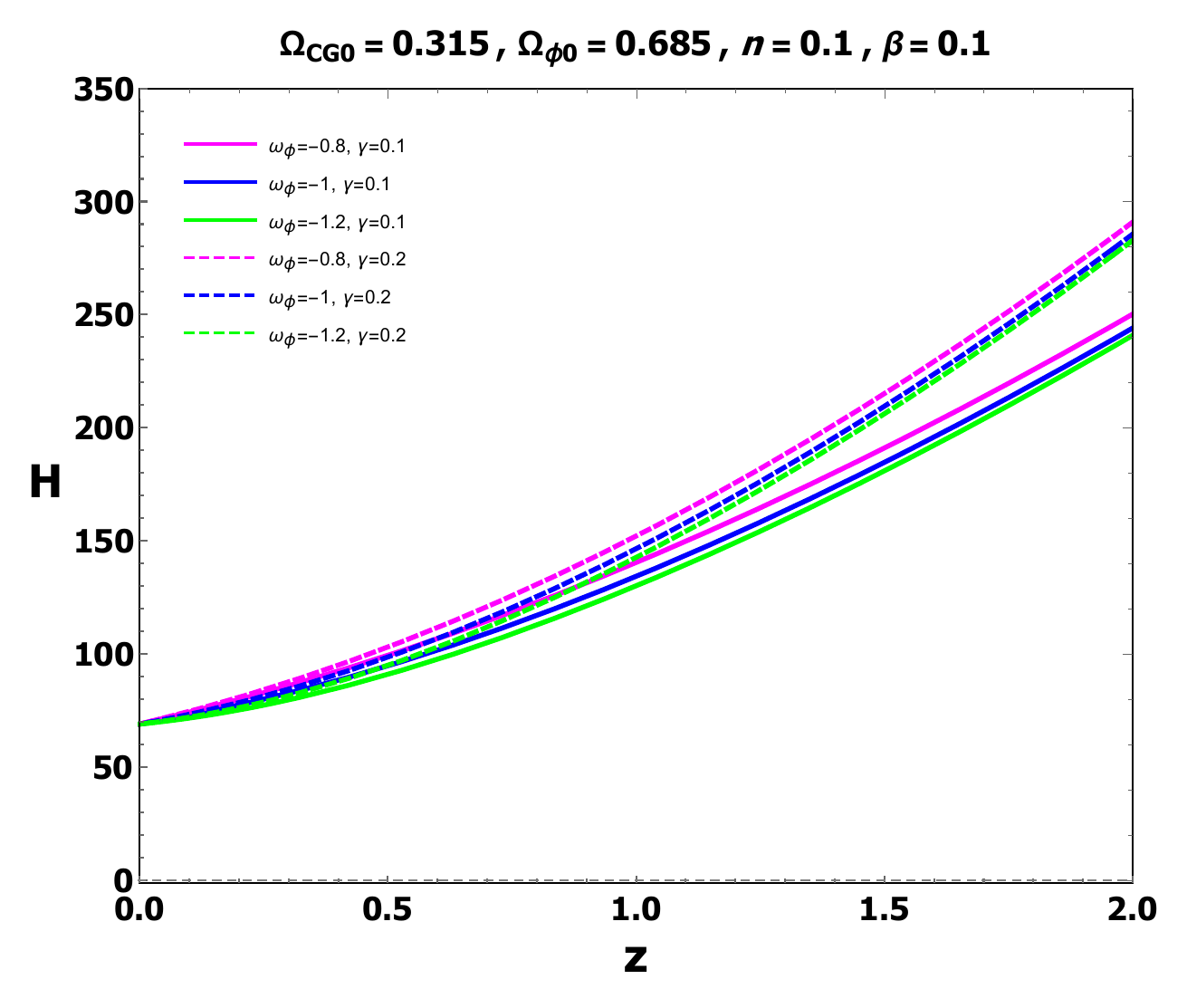}
	
	\caption{The figure show behaviour of Hubble parameter $H$ as a function of redshift z for different values of $\omega_{\phi}$ and $\gamma$. In this figure the present value of the Hubble parameter $H_{0}=69~kms^{-1}Mpc^{-1}$. }
	\label{Hubble}
\end{figure}
Finally, we plot the behavior of the Hubble parameter against the redshift in the figure (\ref{Hubble}) obtained in Eqn (\ref{E}) for some choices of model parameters and the initial values of cosmological parameters: ($n,~\beta$)=(0.1,~0.1) and the cosmological parameters: ($\Omega_{CG0},~\Omega_{\phi0}$)=($0.315,~0.685$) with the present value of the Hubble parameter: $H_{0}=69~kms^{-1}Mpc^{-1}$.



\section{Discussions with concluding remarks}\label{discussion}
We have investigated the cosmological dynamics of two-fluid model of spatially flat FLRW universe. The universe is assumed to be filled with mixture of two non-interacting fluid: one is modified chaplygin gas generalizing the cold matter of the standard $\Lambda$CDM-universe model, and the other a tachyon field generalizing the Lorentz Invariant Vacuum Energy (LIVE) with a constant density represented by the cosmological constant, $\Lambda$, predicted by quantum theory. Tachyon scalar field with inverse square potential ($V\propto \frac{1}{\phi^{2}}$) leads to the four-dimensional autonomous system (\ref{autonomous_system1}) where $\frac{V'(\phi)}{V(\phi) \sqrt{V(\phi)}}$=constant. To achieve the qualitative description of evolutionary dynamics, first we extracted critical points from autonomous system (\ref{autonomous_system1}). To test the stability, we found eigenvalues of linearized Jacobian matrix presented in Table \ref{eigenvalues1}. Local stability is found from eigenvalues of Jacobian matrix for hyperbolic critical points. We employed the numerical computation for finding the stability for non-hyperbolic points. A detailed analysis is presented in section \ref{local_stability}. The corresponding cosmological parameters at the critical points are displayed in the Table \ref{physical_parameters}. From the phase space analysis, we obtained some insightful results which are relevant from cosmological context in early evolution as well as in late-phase. For example:
We obtained some modified chaplygin gas dominated solutions (with $\Omega_{CG}=1,~ \Omega_{\phi}=0$) in the phase space and these do not describe the late-time evolution of the universe. The solutions are represented by the critical points:  $P_1$, $P_2$, $P_{3\pm}$  $P_{4\pm}$, $P_{5\pm}$ and the sets of critical points $P_{8\pm}$, $P_{10\pm}$ and $P_{11}$. In fact, the solutions represented by points $P_1$, $P_{3\pm}$ and the set $P_{8\pm}$ describe the accelerated de Sitter evolution of the universe [$\Omega_{CG}=1$, $\Omega_{\phi}=0$, $\omega_{CG}=-1$, $\omega_{total}=q=-1$] where the the equation of state of MCG behaves as cosmological constant. But, unfortunately, the solutions are unstable in nature in phase space, so, they are unable to provide late-time scenario. These are the transient phase of the evolution. The remaining solutions: $P_2$, $P_{4\pm}$, $P_{5\pm}$, $P_{10\pm}$ and $P_{11}$ are important in early evolutionary era. Some of them ($P_2$, $P_{4\pm}$, $P_{5\pm}$) can describe the radiation ($\omega_{eff}=\frac{1}{3}$) dominated (for $\gamma=\frac{1}{3}$) or dust ($\omega_{eff}=0$) dominated (for $\gamma=0$) decelerated ($q>0$) intermediate phase of the universe where the chaplygin fluid behaves as radiation ($\omega_{CG}=\frac{1}{3}$) or dust ($\omega_{CG}=0$) fluid. Note that the sets $P_{10\pm}$ and $P_{11}$ also describe the dust dominated ($\omega_{eff}=0,~q=\frac{1}{2}$) decelerated intermediate phase where the chaplygin gas behaves as dust ($\omega_{CG}=0$) fluid. \\

We have obtained completely tachyonic fluid dominated solutions (represented by two sets $P_6$ and $P_7$), among which $P_7$ is important in cosmological context, it provides the late-time accelerated evolution of the universe in quintessence era for 
$-\sqrt{\frac{2}{3}}<x_c<\sqrt{\frac{2}{3}},~y_{c}=\sqrt{1-x_{c}^{2}},~ 0<s_{c}<-\frac{1}{3}+y_{c} \sqrt{1-x_{c}^{2}}$ and $n=0$. In this case, the tachyon field behaves as quintessence like fluid.\\

Finally, late-time scaling attractor is achieved by the set of critical points: $P_9$ for all the parameters ($n,~\gamma,~\beta$) related to MCG and for $\lambda=0$ (constant potential) of tachyon scalar field. This depicts the accelerating universe in late-time with similar order of energy densities for dark sectors. This scenario can alleviate the coincidence problem. For a particular choice of $z_c$ the solution can represent chaplygin gas dominated or tachyon field dominated era at late-time. For both the cases, the equation of state for chaplygin and tachyon fluids behave as cosmological constant ($\omega_{CG}=-1,~\omega_{\phi}=-1$). The future evolution will mimic as $\Lambda$CDM. This result is also reflected in the fig. (\ref{Evolution}) where the evolution of the cosmological parameters are plotted and it is confirmed that our model is consistent with the $\Lambda$CDM paradigm. Thus, the dynamical study of the model exhibits different phases of cosmic evolution dominated either by MCG or by Tachyon fluid. For example, the set of points $P_7$ describes late-time accelerated evolution of the universe in quintessence era when it is mostly dominated by tachyon fluid and it behaves as quintessence. Then, set of points $P_9$ represnts the scaling attractor. In this evolution the universe is filled with both the matter: Modified chaplygin gas and the tachyon fluid having similar order of energy densities. It is an interesting result from cosmological point of view. This solution can successfully solve the coincidence problem. Lastly, when the universe is dominated by modified chaplygin gas only, it can not provide any late-time evolution of the universe. Then, the modified chaplygin gas can only be relevant to the early evolution.

Stability of the two fluids (tachyon and MCG) is analysed by evaluating the squared sound speed for the MCG and for tachyon fluid separately. Usually, the classical perturbation of the model is assumed to be stable  when the squared sound speed is positive, i.e., $c_{s}^2 \geq 0$. On the other hand, the ghost instabilities can be realized when $c_{s}^2$ satisfies $c_{s}^2 < 0$ and it violates the causality for $c_{s}^2 > 1$ (i.e., when the sound speed diverges). Then, it is obvious that the condition for stability is $0 \leq c_{s}^2 \leq 1$. Howevever, as discussed in Refs.\cite{Poisson 1995,Lobo 2004} we can not directly say that speed of sound greater than 1 violates the causality as we do not have a speed of sound calculation for the stiff matter. However, we have shown a sufficient condition, and this sufficient condition could in principle provide some phenomenological bound on the model we have considered. We plotted the squared sound speed of tachyonic field for different choices of model parameter $\lambda$ (see in sub-fig. \ref{SS_T}) and  for different choices of model parameters $n,\gamma$ of MCG (in sub-fig.\ref{SS_CG}) and it is found by observing the evolution of the squared sound speed for tachyon and MCG fluids that both the models are stable , i.e., two fluids can be survived at late-times together.\\

We conducted a model independent test for distinguishing our model from other dark energy models by evaluating the statefinder parameters and $O_m$ diagnostic parameters. For that we evaluated the geometrical cosmographic parameters $r$ and $s$ which are obtained from the scale factor $a$. We then investigated the discrimination of our model with $\Lambda$CDM by plotting the trajectories in $r-s$ plane in fig.\ref{sr} and in $r-q$ plane in fig.\ref{qr}. The trajectories in those planes show that our model can mimic the $\Lambda$CDM model at late-times when the EoS for tachyon mimics cosmological constant $\omega_{\phi}=-1$. Further, we study the more general geometric parameters, namely, the jerk ($j$), snap ($s$) and the lerk ($l$) parameters. The behaviour of the trajectories show that the evolution of our model mimics the $\Lambda$CDM in late-times (see in fig.(\ref{CP}) ). The plot of deceleration parameter $q$ (given in Eqn.(\ref{q-transit})) against redshift $z$ in the sub-fig. \ref{zq} exhibits the phase transition of the evolution of the universe at $z=0.32$ from deceleration to acceleration  and there will be an ultimate transition to a $\Lambda$CDM-era at $z=-0.9795$. $O_m$ diagnostic also studied and this shows the model mimics quintessence at past to $\Lambda$CDM in future (see in fig. (\ref{Om})). The behaviour of Hubble parameter is also plotted in the fig. (\ref{Hubble}).

\section*{Acknowledgments}
The authors would like to thank the anonymous reviewers for their insightful and constructive comments so that the paper has improved significantly.
The authors acknowledge the Inter University Centre for Astronomy and Astrophysics (IUCAA), Pune, India where a part of the work has done in an academic visit under IUCAA's visitorship programme. The author Goutam Mandal acknowledges UGC, Govt. of India for providing Senior Research Fellowship [Award Letter No.F.82-1/2018(SA-III)] for Ph.D.

\end{document}